\documentclass[preprint,showpacs,titlepage,aps,prd,
tightenlines,amsmath,byrevtex,nofootinbib]{revtex4}

\usepackage{graphicx}

\def\lsim{\raise0.3ex\hbox{$\;<$\kern-0.75em\raise-1.1ex
\hbox{$\sim\;$}}}
\def\gsim{\raise0.3ex\hbox{$\;>$\kern-0.75em\raise-1.1ex
\hbox{$\sim\;$}}}

\begin{document}

\preprint{hep-ph/0504026}

\title{Resolving Neutrino Mass Hierarchy and CP Degeneracy by \\
Two Identical Detectors with Different Baselines
}


\author{Masaki Ishitsuka$^{1}$}
\email{ishi@suketto.icrr.u-tokyo.ac.jp}
\author{Takaaki Kajita$^{1}$}
\email{kajita@icrr.u-tokyo.ac.jp}
\author{Hisakazu Minakata$^{2}$}
\email{minakata@phys.metro-u.ac.jp}
\author{Hiroshi Nunokawa$^{3}$}
\email{nunokawa@fis.puc-rio.br}
\affiliation{
$^1$Research Center for Cosmic Neutrinos, Institute for Cosmic Ray Research, University of Tokyo, Kashiwa, Chiba 277-8582, Japan \\
$^2$Department of Physics, Tokyo Metropolitan University, Hachioji, Tokyo 192-0397, Japan \\
$^3$Departamento de F\'{\i}sica, Pontif{\'\i}cia Universidade Cat{\'o}lica 
do Rio de Janeiro, C. P. 38071, 22452-970, Rio de Janeiro, Brazil
}

\date{August 6, 2005}

\vglue 1.4cm

\begin{abstract}

We explore the possibility of simultaneous determination of neutrino 
mass hierarchy and the CP violating phase by using two 
identical detectors placed at different baseline distances. 
We focus on a possible experimental setup using neutrino beam 
from J-PARC facility in Japan with beam power of 4MW
and megaton (Mton)-class water Cherenkov detectors, 
one placed in Kamioka and the other at somewhere in Korea. 
We demonstrate, under reasonable assumptions of systematic uncertainties,  
that the two-detector complex with each fiducial volume of 
0.27 Mton has potential of resolving neutrino mass hierarchy up to 
$\sin^2 2\theta_{13} > 0.03$ (0.055) at 2$\sigma$ (3$\sigma$) CL 
for any values of $\delta$ 
and at the same time has the sensitivity to CP violation 
by 4 + 4 years running of $\nu_e$ and $\bar{\nu}_e$ appearance 
measurement. 
The significantly enhanced sensitivity is due to clean detection 
of modulation of neutrino energy spectrum, 
which is enabled by cancellation of systematic uncertainties 
between two identical detectors which receive the neutrino beam 
with the same energy spectrum in the absence of oscillations.  
\end{abstract}

\pacs{14.60.Pq,14.60.Lm,13.15.+g}


\maketitle


\section{introduction}

Despite the great progresses made in varying neutrino oscillation experiments 
\cite{SKatm,solar,KamLAND,K2K,CHOOZ} 
which revealed the structure of the lepton flavor mixing matrix, 
the Maki-Nakagawa-Sakata (MNS) matrix \cite{MNS}, 
there still remains several unanswered questions. 
One of the most intriguing one is the problem of neutrino 
mass hierarchy, a tantalizing dualism of 
the normal ordering and the inverted one where the former (latter) 
implies that a pair of neutrinos with a mass 
gap responsible for solar neutrino oscillations are lighter 
(heavier) than the third neutrino. 
Picking out one from the two alternatives of neutrino mass 
patterns may bring us one of the most significant hints for underlying 
physics of neutrino mass.
Another challenging goal of the next-generation neutrino oscillation 
experiments is to uncover leptonic CP violation. 
It will not only shed light to deep structure behind the complete 
parallelism between quarks and leptons, but also may give us a hint 
for understanding baryon number asymmetry in the universe 
\cite{leptogenesis}.

In this paper, we develop a new experimental strategy of resolving 
the above two unknowns at the same time. 
For the sake of concreteness, we confine ourselves into a general 
framework of the neutrino program supported by 
Japan Proton Accelerator Research 
Complex (J-PARC) located in Tokai village 
\cite{JPARC}.
In particular, we focus on its phase II project with upgraded beam of 4 MW 
and a megaton (Mton) water Cherenkov detector, Hyper-Kamiokande (HK).
The fiducial mass of this detector, in its current design, is 0.54 Mton. 
We propose, instead of placing a 1 Mton detector in the Kamioka mine, 
to break it into two 0.5 Mton detectors\footnote{
A two-detector method for measuring CP violation 
is first described in \cite{MNplb97}.
}
 and place the one in Kamioka 
and the other one at somewhere in Korea.\footnote{
The present proposal may not require significant modification
of the design of HK, because 
the current design of HK consists of two tanks (detectors) located in two 
separated caves due to the constraint on 
the size of the available site \cite {nakamura}.
}
Use of two identical detectors has a definite advantage that 
the systematic errors largely cancel between detectors, which allows 
clean measurement of modulation of neutrino energy spectrum due 
to oscillations.

Principle of determination of mass hierarchy is quite simple.  It may 
be carried out by measuring interference between vacuum and 
matter effects in neutrino oscillations. However, it requires 
sufficiently long baseline of the order of $\sim$ 1000 km. 
Then, the natural possibility one can think of is to place a second 
detector along the beam direction from Tokai to Kamioka 
at somewhere in Eurasian Continent 
\cite{eurasian,MNjhep01,NOVE_mina,SEESAW_hagi}.
Despite many attempts in developing ideas for determining mass 
hierarchy along the line quoted above, no immediately feasible 
proposal has emerged so far.
Probably, the closest one to the goal is the NO$\nu$A project 
(with an upgraded proton driver) in the 
United States \cite{NOVA}.

It is expected in phase II of the J-PARC neutrino program 
that the leptonic CP violating phase $\delta$ can be measured, 
if $\theta_{13}$ is not too small, 
by combining 2 and 6 year measurements of appearance channels 
$\nu_{\mu} \rightarrow \nu_{\rm e}$ and 
$\bar{\nu}_{\mu} \rightarrow \bar{\nu}_{\rm e}$, respectively \cite{JPARC}. 
Though powerful with enormous statistics enabled by intense neutrino 
beam and a huge detector, the measurement of the leptonic CP violating  
phase $\delta$ in the T2K experiment suffers from the obstruction 
due to the unknown mass hierarchy, the fact observed with varying 
degree of robustness and contexts  
\cite{sb-vs-nf,reactorCP,NOON04_yasuda,SEESAW_hagi}. 
In a nutshell, CP violating solution can be confused with CP conserving one 
due to the degeneracy associated with the sign of $\Delta m^2_{31}$ 
\cite{MNjhep01}, i.e., the undetermined mass hierarchy. 
Thus, the problems of measuring $\delta$ and determining mass 
hierarchy inherently couple with each other through the parameter 
degeneracy.

In this paper we explore a completely different strategy 
from these previous attempts. 
The tightly coupled feature of problems of determining mass hierarchy 
and $\delta$ naturally suggests 
a new radical approach of simultaneous determination of these two quantities. 
For this purpose, we focus on a two detector system, 
one detector in Kamioka and another one 
at somewhere in Korean peninsula.

The T2K experiment will use an off-axis beam, where the direction
of the charged pions (with some contamination of kaons and muons)
is 2 to 3 degrees different from the direction to the neutrino detector from 
the target. By selecting an angle between the directions of the charged 
particles and the neutrinos (off-axis angle) low energy neutrino beam 
whose maximum energy ranges from about 500~MeV (for 3 degree 
off-axis angle) to 800~MeV (for 2 degree off-axis angle) can be produced.
Because of the curvature of the surface of the earth, the direction to Kamioka,
which is 295~km away from the target, is 1.3 degrees below the horizon as 
seen from the neutrino production target. 
According to the design of the T2K beam-line,
the 2-to-3-degree off-axis beam is realized by 
directing the charged pion beam by 1.0 degree north of HK
and 3.0 to 4.1 degrees below the horizon. The Korean peninsula is about 1000
to 1250~km away from the J-PARC facility. Therefore, the direction
to the Korean peninsula is 4.5 to 5.6 degrees below the horizon.
With this beam-line configuration, we find that the off-axis beam with the 
off-axis angle larger than 1~degree is automatically available in South 
Korea \cite{SEESAW_hagi} for the 2.5 degree off-axis beam in Kamioka. 
We, therefore, find that the  
2.5 degree (or more generally any angle between 2 and 3 degrees)
off-axis beam in Kamioka and Korea is simultaneously 
available with the present
T2K beam-line configuration. Later we find that the identical beam 
spectrum at two different positions is very powerful in determining 
the mass hierarchy.

In Sec.~\ref{strategy}, we develop our strategy of how to resolve 
the mass hierarchy and CP degeneracies. 
In Sec.~\ref{degeneracy}, we present
a pedagogical discussion of the parameter degeneracy to 
a minimal level for reader's understanding, and then illustrate 
how it can be resolved by spectral informations. 
Those who are familiar to the phenomenon should skip this section.
In Sec.~\ref{comparison}, we present quantitative analysis of 
sensitivity of resolving the parameter degeneracy and 
determining neutrino mass hierarchy. 
In Sec.~\ref{conclusion}, we give concluding remarks.


\section{Right and wrong ways of using the Korean Detector}
\label{strategy}

\subsection{A detector in Korea is not always powerful}

\begin{figure}[htbp]
\vglue 0.3cm
\begin{center}
\includegraphics[width=0.96\textwidth]{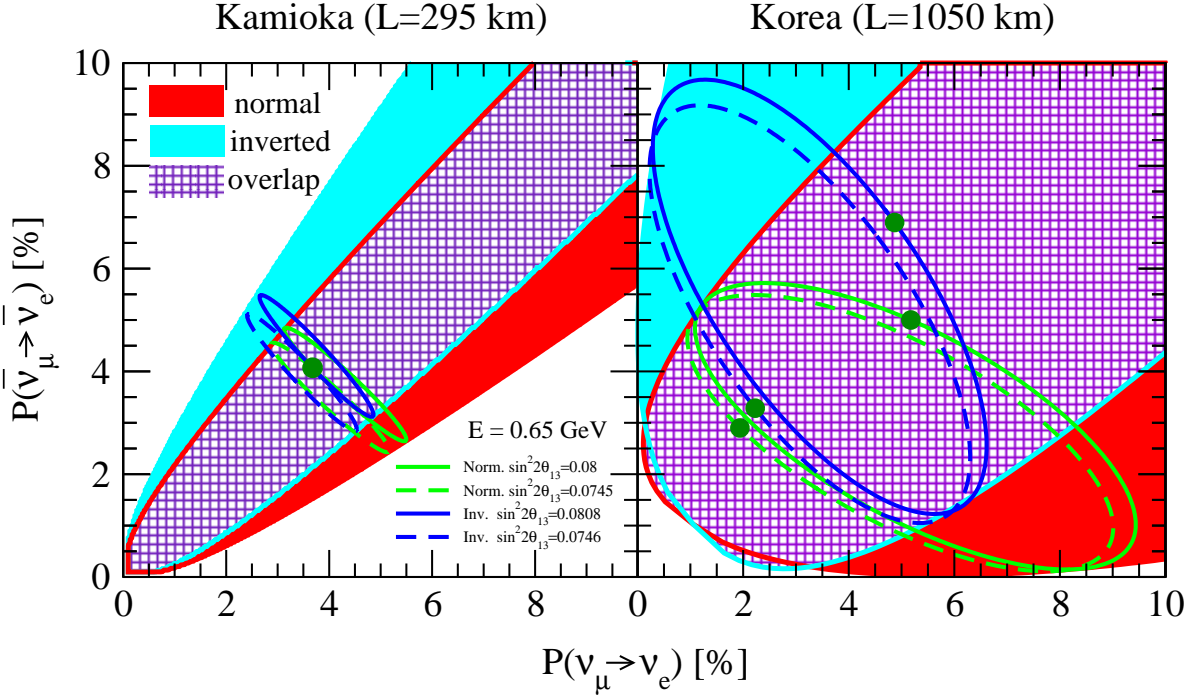}
\end{center}
\vglue -0.7cm
\caption{ Regions spanned by neutrino and 
anti-neutrino appearance probabilities 
$P(\nu_{\mu} \rightarrow \nu_{\rm e})$ and 
$P(\bar{\nu}_{\mu} \rightarrow \bar{\nu}_{\rm e})$
in a bi-probability space when CP phase $\delta$ and $\theta_{13}$ 
are varied. 
Neutrino energy is taken as 650~MeV, and the baseline distances to 
Kamioka and Korea detectors are 295 km and 1050 km, respectively. 
The two regions with dark-gray (red) and light-gray (blue) 
correspond to the normal ($\Delta m^2_{31}>0$) and 
inverted ($\Delta m^2_{31}<0$) mass hierarchies, respectively.   
The left and right panels are for a detector in Kamioka and 
in Korea, respectively.
A single filled circle in the left panel corresponds to four degenerate 
solutions allowed by the measurement in Kamioka, and they 
split into four different points in the right panel, which may be 
resolved by the Korean detector.
Throughout this paper, mixing parameters other 
than $\theta_{13}$ and $\delta$ 
are fixed to be $\Delta m^2_{21} = 7.9 \times 10^{-5}$ eV$^2$, 
$\sin^2 \theta_{12} = 0.29$, 
$\Delta m^2_{31} = \pm 2.5 \times 10^{-3}$ eV$^2$ and 
$\sin^2 2\theta_{23} = 1$, unless otherwise stated.
Matter densities along the neutrino trajectories 
are assumed to be constant as 2.3 and 2.8 g/cm$^3$ for
Kamioka and Korea, respectively.}
\label{biP1}
\end{figure}

We start by giving a cautionary remark. 
One must be aware that a detector in Korea which 
receives the same neutrino beam from J-PARC facility is {\it not} 
immediately a good idea for detecting the matter effect and thereby 
determining the neutrino mass hierarchy.
To understand the point, we draw in Fig.~\ref{biP1} regions 
spanned by probabilities of neutrino and 
anti-neutrino appearance probabilities 
$P(\nu_{\mu} \rightarrow \nu_{\rm e})$ and 
$P(\bar{\nu}_{\mu} \rightarrow \bar{\nu}_{\rm e})$
in a bi-probability space \cite{MNjhep01}.
For a given $\theta_{13}$ variation of $\delta$ from 0 to $2\pi$ 
gives an ellipse and it forms a region when $\theta_{13}$ is varied.
Left and right plots in Fig.~\ref{biP1} are, respectively, for 
a detector in Kamioka and a one in Korea.
(See \cite{MNjhep01} for a detailed explanation of properties of 
the bi-probability plot.)

The two regions with dark-gray (red) and light-gray (blue) 
correspond to the normal ($\Delta m^2_{31}>0$) and 
inverted ($\Delta m^2_{31}<0$) mass hierarchies, respectively. 
Here, we define the mass squared difference of neutrinos as 
$\Delta m^2_{ji} \equiv m^2_j - m^2_i$ where
$m_i$ is the eigenvalue of the $i$th mass-eigenstate.
As one can see, the fraction of  overlap between the two regions 
in the Korean detector is as large as the one in Kamioka, 
indicating that there exists confusion of mass hierarchy 
in Korea as heavy as that in Kamioka.
If one relies on the usual method of comparing 
$P(\nu_{\mu} \rightarrow \nu_{\rm e})$ to 
$P(\bar{\nu}_{\mu} \rightarrow \bar{\nu}_{\rm e})$, 
the Korean detector is as bad as that in Kamioka. 
Of course, it implies that overall the Korean detector
is worse because the number of events is an 
order of magnitude smaller for a given flux.

The behavior we just saw above can be understood 
qualitatively by looking into 
the appearance oscillation probability 
$P(\nu_{\mu} \rightarrow \nu_{\rm e})$ (and its charge-conjugation) 
valid to first-order in matter effect \cite{AKS}, 
\begin{eqnarray}
P[\nu_{\mu}(\bar{\nu}_{\mu})
&\rightarrow& \nu_{\rm e}(\bar{\nu}_e)] 
\nonumber \\
&=&
\sin^2{2\theta_{13}} s^2_{23}
\sin^2 \left(\frac{\Delta m^2_{31} L}{4 E}\right)
-\frac {1}{2}
s^2_{12}\sin^2{2\theta_{13}}s^2_{23}
\left(\frac{\Delta m^2_{21} L}{2 E}\right)
\sin \left(\frac{\Delta m^2_{31} L}{2 E}\right) 
\nonumber \\
&+&
2J_{r} \cos{\delta}
\left(\frac{\Delta m^2_{21} L}{2 E} \right)
\sin \left(\frac{\Delta m^2_{31} L}{2 E}\right) \mp 
4J_{r}\sin{\delta}
\left(\frac{\Delta m^2_{21} L}{2 E}\right)
\sin^2 \left(\frac{\Delta m^2_{31} L}{4 E}\right) 
\nonumber\\
&\pm& \cos{2\theta_{13}}
\sin^2{2\theta_{13}} s^2_{23}
\left(\frac {4 Ea(x)}{\Delta m^2_{31}}\right)
\sin^2 {\left(\frac{\Delta m^2_{31} L}{4 E}\right)}
\nonumber \\
&\mp&
\frac{a(x)L}{2}\sin^2{2\theta_{13}}\cos{2\theta_{13}} s^2_{23}
\sin \left(\frac{\Delta m^2_{31} L}{2 E}\right) + 
c^2_{23} \sin^2{2\theta_{12}} 
\left(\frac{\Delta m^2_{21} L}{4 E}\right)^2 
\label{Pmue}
\end{eqnarray}
In (\ref{Pmue}), $a(x)= \sqrt 2 G_F N_e(x)$  \cite{wolfenstein} 
where $G_F$ is the Fermi constant, $N_e(x)$ denotes the
electron number density at $x$ in the earth,
$J_r$ $(= c_{12} s_{12} c_{13}^2 s_{13} c_{23} s_{23} )$ 
denotes the reduced Jarlskog factor, $E$ is the neutrino energy, 
$L$ is the baseline distance, 
and the upper and lower sign $\pm$ refer to the neutrino and 
anti-neutrino channels, respectively. 
We have added a term due to solar mass scale oscillations, 
the last term in (\ref{Pmue}), whose value of order 0.01 at 
a Korean detector is non-negligible.  
One can easily observe that when we move from the first to the 
second oscillation maxima, the size of CP phase effects become 
larger by a factor of 3 while the matter effect stays the same. 
At off the oscillation maxima the first term in last line of (\ref{Pmue}) 
makes additional contributions which render the fraction of matter 
effect larger, but only up to a level given in Fig.~\ref{biP1} 
due to a modest size of the matter effect, 
$aL=0.54 (\rho/2.8\ \mbox{g\ cm}^{-3})(L/1000\ \mbox{km})$. 
Therefore, a second detector at the 
second oscillation maximum, though attractive because of a 
factor of 3 larger effect of CP phase $\delta$ is not the best 
place to discriminate the mass hierarchy \cite{NOVE_mina}.

In summary, it appears that there is no obvious merit of placing a 
detector in Korea to measure the matter effect. 
To really utilize the attractive feature of a factor of 3 larger CP effect 
we must go beyond the level of our above discussions, 
as we will pursue in the next subsections.

\subsection{Energy dependence is far more dynamic if seen by the 
Korean detector}

To uncover possible advantages of the Korean detector we have 
examined how the appearance probabilities depend upon the energy 
and baseline. We observed that after neutrinos pass through 
the second oscillation maximum the appearance probabilities 
sharply fall. The behavior, together with an enhanced matter effect, 
produces a dynamic behavior of the energy spectrum of oscillated neutrinos. 
The features of the oscillation probabilities suggest the baselines 
between the second oscillation maximum and the subsequent minimum, 
between about 900 and 1200 km, as appropriate ones. 
It is an accidental coincidence that the mountainous area with which 
neutrino beam from J-PARC facility first encounter is located at 
about 1000-1100~km away from Tokai village. 
Therefore, we take the baseline distance of 1050 km as a typical 
distance to the Korean detector throughout our analysis in this paper.
Noting that 
$\Delta m^2_{31} L/2 E = 3.11 \pi
(\Delta m^2_{31}/2.5 \times 10^{-3}\ \text{eV}^2) (L/1000\ \mbox{km})
(E/0.65\ \text{GeV})^{-1}$, $L=1050$ km corresponds to 
about 10\% off the second oscillation maximum at peak energy. 
The choice matches with our earlier observation that 
it is better to stay somewhat away from the second oscillation maximum 
to enhance the matter effect so that the last term in Eq. (\ref{Pmue})
makes some contributions to the probabilities.

\begin{figure}[htbp]
\vglue 0.3cm
\begin{center}
\includegraphics[width=0.96\textwidth]{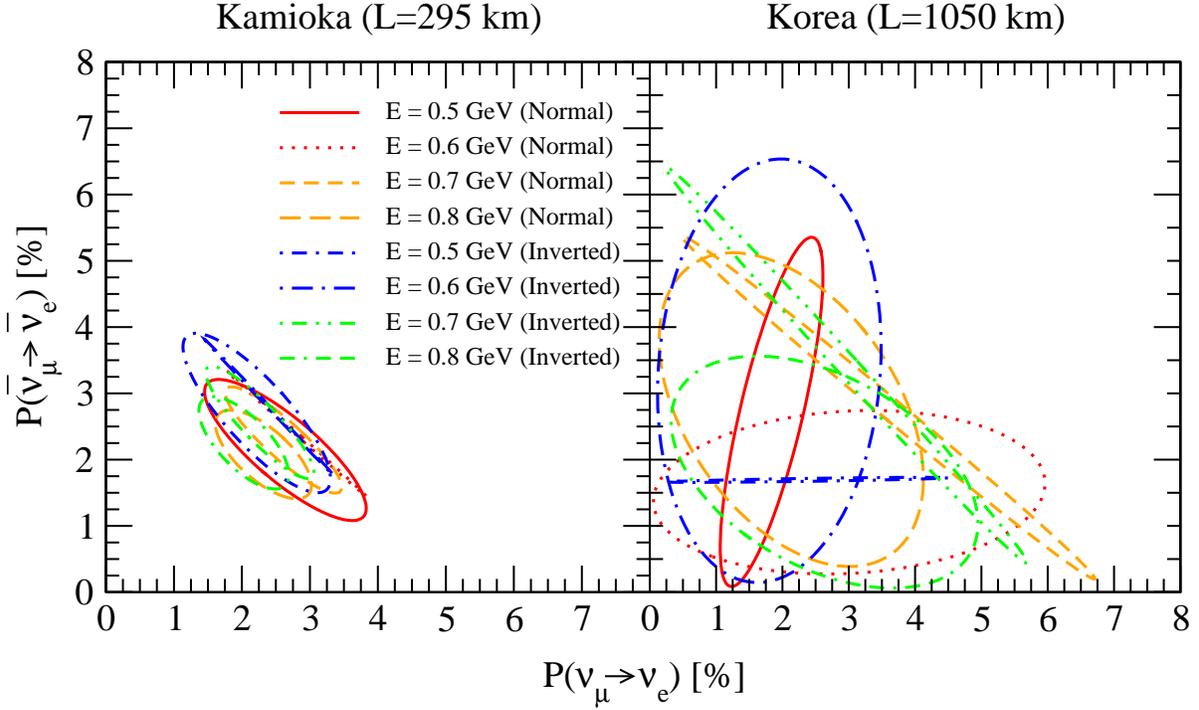}
\vglue -0.3cm
\end{center}
\vglue -0.5cm
\caption{Energy dependences of the oscillation probabilities 
for $\sin^2 2\theta_{13}=0.05$ are represented by plotting ellipses 
(which results as $\delta$ is varied from 0 to $2\pi$) 
in bi-probability space for various neutrino energies from 0.5 to 0.8 GeV.
Other mixing parameters are fixed as in the caption of Fig.\ref{biP1}.
The left and the right panels are for detectors in Kamioka and in 
Korea, respectively.
The correspondence between energies and the line symbols 
are denoted in the figure; 
The ellipses in upper 4 symbols (warm colors) indicate the ones of 
normal mass hierarchy ($\Delta m^2_{31} >0$)
and the one of lower 4 symbols (cold colors) the ones of 
inverted mass hierarchy ($\Delta m^2_{31} <0$). 
}
\label{biP2}
\end{figure}

To demonstrate the dynamical behavior of the oscillated neutrino spectrum, 
we draw another bi-probability plot in Fig.~\ref{biP2} to represent 
how different are the energy dependence of the oscillation 
probabilities when they are seen by detectors in Kamioka and in Korea. 
We notice the striking difference between 
the left and right panels of Fig.~\ref{biP2}; 
In contrast to the quiet behavior in Kamioka, 
the ellipses are larger in size and are extremely dynamic in 
movement in Korea, when the energy is varied from 0.5 to 0.8 GeV.

Let us summarize the notable differences between Kamioka and Korea, 
because they are all crucially important as the building blocks of our strategy. 

\begin{itemize}

\item

The variation of the appearance probabilities due to $\delta$
is larger by a factor of 2-3 in Korea compared to that in Kamioka. 
It is notable that the ellipses with energies from 0.5 to 0.8 GeV span 
almost entire area of triangular region of appearance probabilities less than 0.07.
It is partly due to about a factor 3 larger CP effect, and to the 
enhanced solar term which scales approximately as $L^2$
(see Eq. (\ref{Pmue})).

\item

The energy dependence of the ellipse is very dramatic in the Korean detector. 
When the energy is varied from 0.5 to 0.8 GeV, the slope 
of the major axis moves counterclockwise (clockwise) 
by more than $\pi/2$ in the case of normal (inverted) hierarchy, 
which is represented with the symbol exhibited in Fig.~\ref{biP2}. 
(For electronic version,  the ellipses of the normal and the inverted 
hierarchies are represented by warm and cold colored lines, respectively.) 
It implies that CP phase $\delta$ affects differently onto 
different part of the energy spectrum. 
Notice that positive slope of  the major axis has never been 
observed in region around the first oscillation maximum. 
In contrast to such violent behavior the energy dependence 
of the ellipse is very quiet in Kamioka.

\end{itemize}

\noindent
This strong energy dependence of the appearance oscillation 
probabilities in Korea, which are different for normal and 
inverted hierarchies and are 
helped by their enhanced size, is the key to the 
enormous sensitivity of the two-detector complex as we will fully 
explore in Sec.~\ref{comparison}.

We want to mention here about the BNL proposal in which a 
wide-band beam from upgraded AGS will be detected by the far 
detector located at 2950 km to observe the effects of multiple 
(1-3) oscillation maxima \cite{BNL}. 
The dramatic difference in the energy dependence of the neutrino 
oscillation probability between the first and the second oscillation 
maxima may have been contributed to the high sensitivity 
(despite such a long baseline) obtained in the proposal.

\subsection{Determination of mass hierarchy through resolving degeneracy}

Now we have reached to the point to establish our new strategy 
for the determination of the neutrino mass hierarchy. 
In a nutshell we do it by resolving the parameter degeneracy 
associated with the unknown sign of $\Delta m^2_{31}$.
While we postpone an organized discussion of the parameter 
degeneracy to Sec.~\ref{degeneracy}, it will be shown there that 
to resolve the $\Delta m^2_{31}$-sign degeneracy, the spectrum shape 
analysis of the appearance events in two detectors in different baseline 
(or at different energy) is crucial. 
Then, by placing the two detectors in Kamioka and in Korea, 
we can fully utilize the remarkable difference in energy dependence 
of the oscillation probabilities to solve the parameter degeneracy.

There exists great merit of using two identical detectors together 
with the neutrino beam of identical spectrum shape 
(in the absence of oscillations) which 
makes demonstration of the CP violation and resolution of the 
mass hierarchy degeneracy much cleaner.\footnote{
We are not aware to whom the idea of error cancellation with use of 
identical detectors can be attributed. 
It was utilized, for example, in the Bugey experiment \cite{bugey}, 
and there were many examples of constructing near detectors 
more or less identical with far detectors.
}
We will also fully utilize this characteristics of Kamioka-Korea two
identical detector
complex in our analysis in Sec.~\ref{comparison}.

\section{Parameter degeneracy and its resolution by spectral analysis}
\label{degeneracy}

In this section, we briefly summarize the parameter degeneracy in 
neutrino oscillations \cite{Burguet-C,MNjhep01,octant,BMW1,KMN02,MNP2} 
to the extent that is necessary to understand the rest of this paper. 
We then illustrate how it can be resolved with use of the Kamioka-Korea 
two identical detectors, which greatly differ in energy dependences 
they observe. 
Those who are familiar to the parameter degeneracy should skip 
this section and directly go to Sec.~\ref{comparison}, where full details of 
the quantitative analysis of resolving power of the degeneracy 
will be presented.
%

It should be obvious from Fig.~\ref{biP1} that a set of measurement 
of $P(\nu_{\mu} \rightarrow \nu_e)$ and 
$P(\bar{\nu}_{\mu} \rightarrow \bar{\nu}_e)$ 
at a particular baseline and an energy allows four solutions 
(crossing points of the ellipses) 
of $\theta_{13}$ and $\delta$.
The origin of the four-fold solutions may be understood as 
the intrinsic degeneracy of $\theta_{13}$ and $\delta$ \cite{Burguet-C} 
(a single crossing point allows two ellipses), 
duplicated by the unknown 
sign of $\Delta m^2_{31}$ \cite{MNjhep01} which admits opposite 
$\Delta m^2_{31}$-sign solutions.
The mixed $\Delta m^2_{31}$-sign degenerate solution occurs in 
the cross-hatched region in Fig.~\ref{biP1}.
A degenerate solution in Kamioka (left panel) 
splits into four separate points (right panel) in the Korean detector.
The importance of the Korean detector is clearly understood.

In fact, the four-fold degeneracy is further multiplied by the octant 
ambiguity of $\theta_{23}$ for a given $\sin{2\theta_{23}}$ \cite{octant}, 
leading to the total eight-fold degeneracy  \cite{BMW1}.
But we do not try to resolve the $\theta_{23}$ degeneracy in this paper 
by simply assuming that $\theta_{23}=\pi/4$ in most part of our analysis.
We, however, examine the effect of non-maximal $\theta_{23}$ on 
resolution of neutrino mass hierarchy and CP degeneracy.\footnote{
%
It is well recognized that it is very difficult to resolve the degeneracy 
associated with $\theta_{23}$ if we rely only on accelerator measurement 
using $\nu_{\mu}$ beam. 
It in turn implies that our discussion in this paper may be stable 
against inclusion of deviation of $\theta_{23}$ from the maximal. 
Lifting the $\theta_{23}$ degeneracy may be 
done only when the accelerator measurement of 
$\nu_{\mu} \rightarrow \nu_{\rm e}$ is combined with either pure 
measurement of $\theta_{13}$ by reactor experiments \cite{MSYIS}, 
or the complementary information from the silver channel, 
$\nu_{e} \rightarrow \nu_{\tau}$~\cite{silver}.
}

\begin{figure}[htbp]
\begin{center}
\hspace{-2.0mm}
\includegraphics[width=0.79\textwidth]{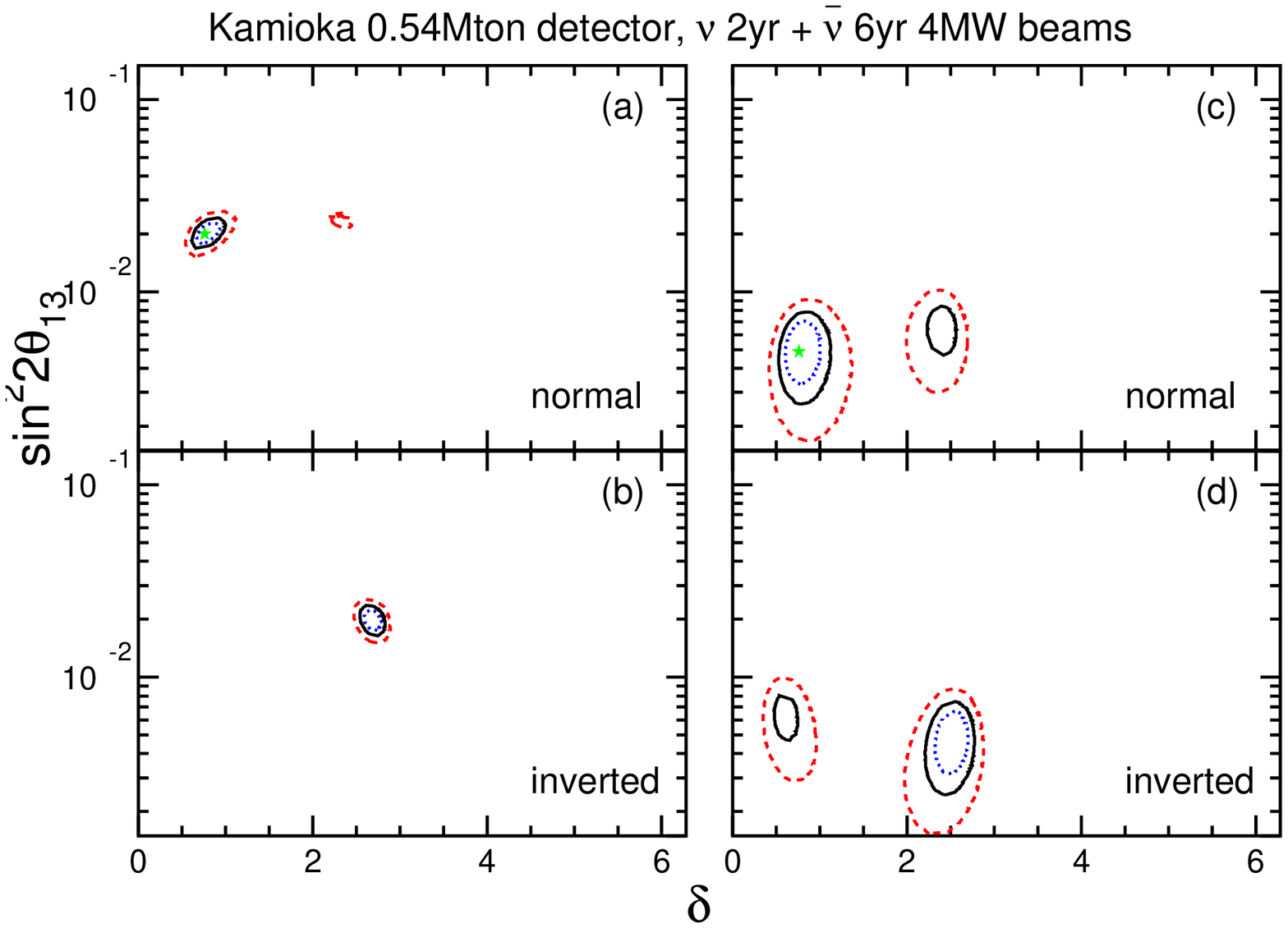}
\includegraphics[width=0.80\textwidth]{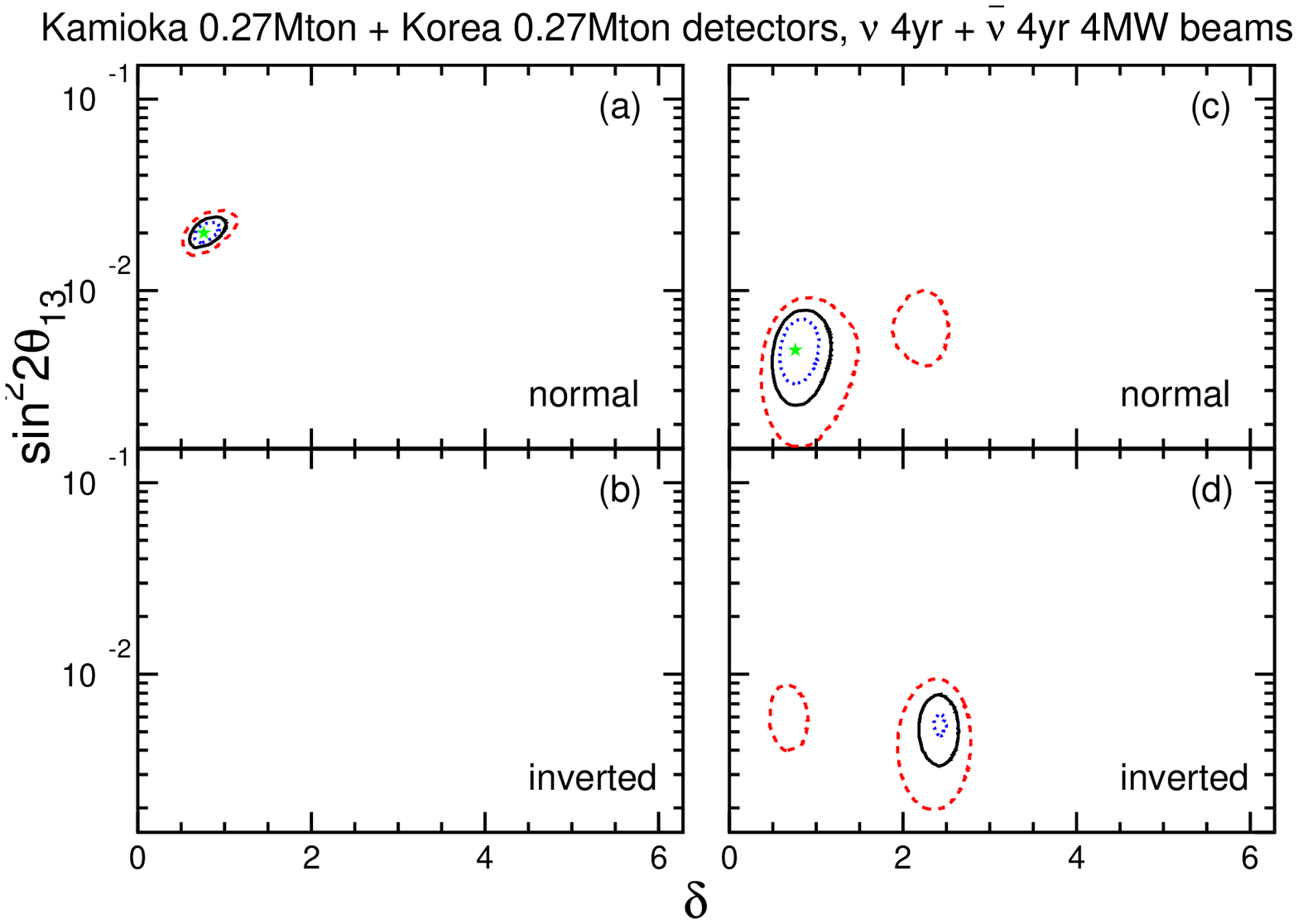}
\end{center}
\vglue -0.7cm
\caption{Examples of the parameter degeneracy.
The true solutions are assumed to be located at 
($\sin^2{2\theta_{13}}$ and $\delta$) = (0.005, $\pi/4$)
(left panels)
and (0.02, $\pi/4$) (right panels) 
with positive sign of $\Delta m_{31}^2$, 
as indicated as (green) stars. Three contours in each figure correspond to
the 68\% (dotted lines, blue), 90\% (solid lines, black) and 99\% 
(dashed lines, red) C.L. sensitivities, which are defined
as the difference of the $\chi^2$ being 2.30, 4.61 and 9.21, respectively.
The upper and lower panels are for a detector in Kamioka
and for two detectors one in Kamioka and the other in Korea 
(twin HK), respectively. 
}
\label{4fold}
\end{figure}

Now we present in Fig.~\ref{4fold} a four-fold degenerate solutions 
to show how it looks like in the plane spanned by $\sin^2{2\theta_{13}}$ 
and $\delta$ obtained as a result of our analysis whose details will be fully 
explained in Sec.~\ref{comparison}.
These solutions are obtained by assuming the operation of the J-PARC 
experiment for total 8 years running with $\nu_{\rm e}$ and $\bar{\nu}_{\rm e}$ 
appearance channels. 
The upper figure: 2 years running of the neutrino mode and 6 years running of 
anti-neutrino mode with single detector of 0.54 Mton of the fiducial volume 
placed in Kamioka only.
The lower figure: 4 years of running of both in neutrino and anti-neutrino 
modes with two detectors of 0.27 Mton of the fiducial volume placed one in 
Kamioka and the other in Korea.
In each figures  the left and the right panels are for cases with 
true values of $\theta_{13}$ given by $\sin^2{2\theta_{13}}=0.02$ and 
$\sin^2{2\theta_{13}}=0.005$, respectively, with 
positive value of $\Delta m_{31}^2$, as inputs.

Let us understand the features of the four solutions represented in 
Fig.~\ref{4fold}. First, we focus on the right panel of Fig.~\ref{4fold}.
We observe two distinct pairs of solutions, the same 
$\Delta m^2_{31}$-sign (intrinsic) degeneracy 
and the mixed $\Delta m^2_{31}$-sign degeneracy.

\begin{itemize}

\item

{\bf the same $\Delta m^2_{31}$-sign degeneracy}

\end{itemize}

One can identify a pair of the solutions in each mass 
hierarchy, which obviously corresponds to the intrinsic degeneracy. 
It is known that the two solutions of $\delta$ are approximately 
symmetric with respect to reflection at $\delta=\pi/2$, that is 
$\delta_{2} - (\pi - \delta_{1}) \lsim 0.02\pi$ \cite{MNP2,nufact03_mina}. 
Within the same mass hierarchy, the transformation 
$\delta \rightarrow (\pi - \delta)$ produces flipping the sign of 
$\cos{\delta}$ term in the appearance probability in Eq.($\ref{Pmue}$).
Therefore, 
the oscillation probabilities with these two degenerate solutions 
have different energy dependence, and hence they are discriminable 
by doing spectrum analysis.
This is what happens, though incompletely, in the case of 
$\sin^2{2\theta_{13}}=0.005$ (right panels).
For $\sin^2{2\theta_{13}}=0.02$ (left panels) the spectral information is powerful 
enough to resolve (almost) the intrinsic degeneracy even for a single 
detector in Kamioka, as seen in the left panels of the upper plots 
in Fig.~\ref{4fold}.
Use of the spectral informations to solve the parameter degeneracies 
has been discussed by many authors; 
See for example  \cite{Burguet-C,MNjhep01,sb-vs-nf,spectral_resol}.

\begin{figure}[htbp]
\begin{center}
\includegraphics[width=1.0\textwidth]{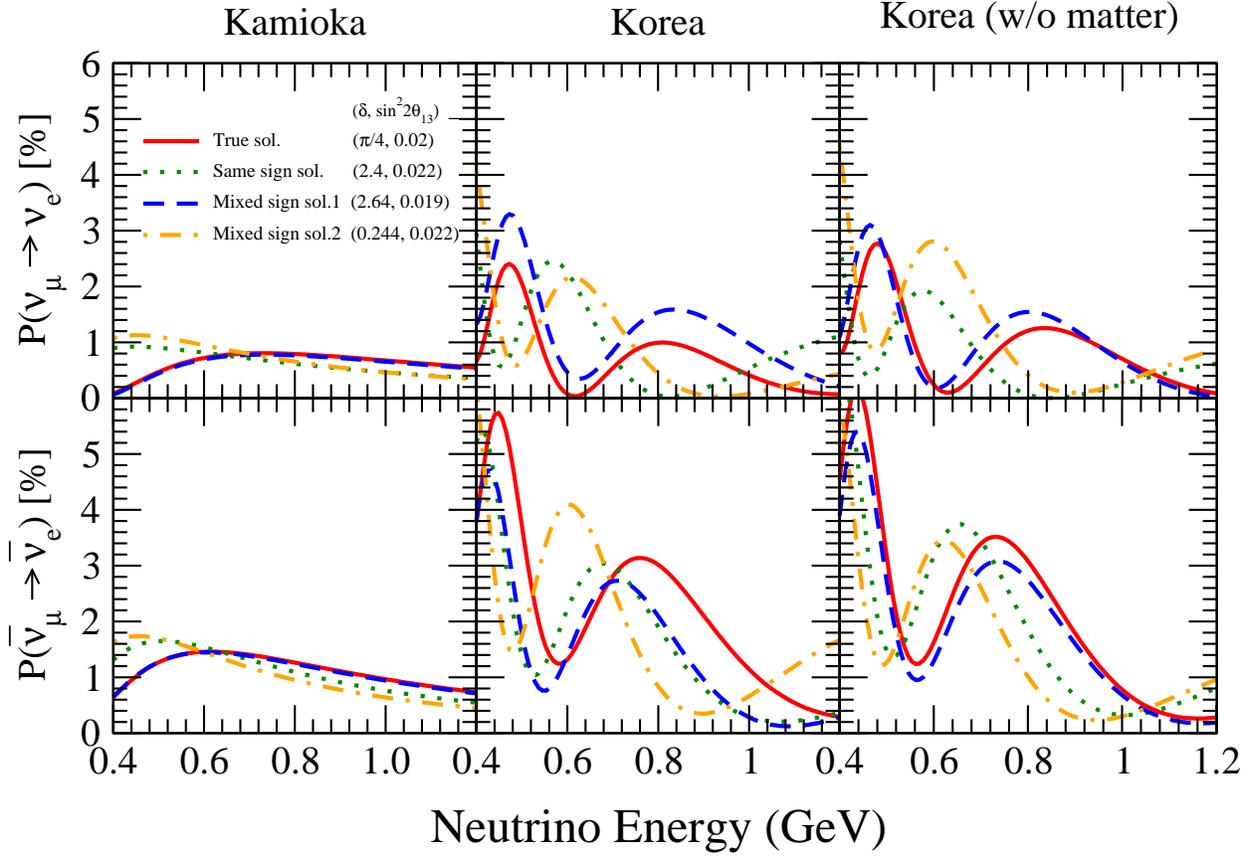}
\end{center}
\vglue -0.5cm
\caption{Neutrino oscillation probabilities corresponding to a 
four-fold degenerate solutions obtained by measurement in 
Kamioka by the rate only analysis are plotted as a function of 
neutrino energy. 
Left panels: appearance probabilities in Kamioka. 
Middle panels: appearance probabilities in Korea.
Right panels: appearance probabilities in Korea, but without the matter
effect.
}
\label{spectral}
\end{figure}

\begin{itemize}

\item

{\bf the mixed $\Delta m^2_{31}$-sign degeneracy}

\end{itemize}

One can also identify another pair of solutions which are still allowed 
even at 68\% CL in the opposite-sign $\Delta m^2_{31}$ panels. 
This is the degenerate solution due to the unknown sign of 
$\Delta m^2_{31}$ (or type of mass hierarchy). 
It arises because of the approximate symmetry under the 
simultaneous transformation \cite{MNjhep01} 
(which applies if the matter effect is small) 
\begin{eqnarray}
& &\delta \rightarrow \pi - \delta
\hskip 0.5 cm (\mbox{mod.} 2 \pi),
\nonumber \\
& &\Delta m^2_{31} \rightarrow - \Delta m^2_{31}.
\label{flipsym}
\end{eqnarray}
Since it is valid apart from tiny corrections of the order of 
$\sin^2{2\theta_{13}} (\Delta m^2_{21}/\Delta m^2_{31})$ 
these two solutions 
{\it cannot} be distinguished only by spectrum analysis at the 
single detector with small matter effect. 
It is the reason why we need a Korean detector to resolve the 
degeneracy due to mass hierarchy. 
The fact that the same transformation of $\delta$ appears in 
(\ref{flipsym}) explains the reason why two pairs of the 
mixed $\Delta m^2_{31}$-sign solutions forms X (cross) shape 
in Fig.~\ref{4fold}.

\vspace{0.2cm}

To have a feeling on how the spectrum information helps to resolve 
the degeneracies, we present in Fig.~\ref{spectral} the energy 
dependences of the oscillation probabilities 
to be observed at detectors in 
Kamioka (left panels) and in Korea (middle and right panels). 
In the right panels, we artificially switch off the matter effect to illuminate 
its role in resolving the degeneracy.
We pick up a set of mixing parameters which correspond to a degenerate 
solution in Kamioka from the rate only analysis, which are shown 
in the figure, and indicate how these four solutions differ in 
energy dependences in Kamioka and Korea. 
The value taken for $\theta_{13}$, $\sin^2{2\theta_{13}} \simeq 0.02$, 
is large enough for the Kamioka-Korea two-detector complex to solve 
all the degeneracies, but is too small for a single detector in Kamioka 
to do the job, as indicated in the left panels of Fig.~\ref{4fold}. 
It is obvious from Fig.~\ref{spectral} that the Korean detector 
sees very different energy spectrum of neutrinos. 
We also note the several characteristic features:

\vspace{0.2cm}

\noindent
(1) In Kamioka (left panels), there are some differences between 
the same-sign degeneracy and the mixed-sign degeneracy. 
In the former case energy dependences of a pair of the solutions 
are somewhat different, leading to resolution of  intrinsic degeneracy 
(as is seen in the left panels of Fig.~\ref{4fold}) 
if the statistics is sufficient and the systematic error is controlled. 

\vspace{0.2cm}

\noindent
(2) In Korea (middle panels), the energy dependences of the four solutions 
are so distinct, and it is conceivable that the spectral information 
solves all the four-fold degeneracies, provided that the statistics 
is sufficiently large.

\vspace{0.2cm}

\noindent
(3) Comparison between the middle and the right panels of Fig.~\ref{spectral}
reveals that the matter effect plays a key role in resolving the 
$\Delta m^2_{31}$-sign degeneracy (see solid red and blue dashed curves).

\section{Sensitivities of resolving mass hierarchy and CP 
degeneracy}
\label{comparison}

In the first subsection we give a full details of our analysis, including 
treatment of experimental errors, treatment of background, and 
the statistical procedure which is used to investigate the sensitivity of 
resolving the degeneracies.
In the next subsections we give our results and perform 
stability checks of the results against the uncertainty of the matter density, 
the change in the systematic errors,  
and the possible deviation of $\theta_{23}$ from the maximal value.

\subsection{Assumptions and the definition of $\chi^2$}
\label{chi2}

In order to understand the advantage of the two detector system at 
295~km (Kamioka) and 1050~km (Korea), we carry out 
a detailed $\chi^2$ analysis. 
We use various numbers and distributions available from
references related to T2K \cite{JPARC,JPARC-detail}.
Here we summarize the main assumptions and the methods 
used in the $\chi^2$ analysis.

We use the reconstructed neutrino energy for single-Cherenkov-ring 
electron-like events
(about 80\% of which are quasi-elastic charged-current (CC) $\nu_e$ interactions).
The resolution in the reconstructed neutrino energy is 80~MeV 
for quasi-elastic events. 
We assume that $\sin^2 2\theta_{23}$ and 
$|\Delta m_{31}^2|$ should be known precisely by the time when 
the experiments we consider in this report will be carried out. 
We take for these parameters 
$\sin^2 2\theta_{23}=1$ and 
$\Delta m_{31}^2 = \pm 2.5 \times 10^{-3} \text{eV}^2$. 
Hence, we assume that the 
energy spectrum of the beam is the one expected by the
2.5~degree off-axis-beam in T2K. The shape of the
energy spectrum for the anti-neutrino beam is assumed to be
identical to that of the neutrino beam. The event rate
for the anti-neutrino beam 
in the absence of neutrino oscillations
is smaller by a factor of 3.4 due mostly to the lower neutrino 
interaction cross sections and partly to the slightly lower 
neutrino flux.

In the T2K experiment, 
28 background events are expected 
for the reconstructed neutrino energies between 350 and 850~MeV for
$(0.75 \times 0.0225 \times 5) \text{MW} \cdot \text{Mton} \cdot \text{yr}$ 
measurement 
with the neutrino beam. The energy dependence 
of the background rate and the rate itself are taken from \cite{JPARC-detail}.
The background rate is expected to be higher in the lower neutrino energies.
The expected number of signal events
is 122 for $\sin^2 2\theta_{13} =$0.1 with the same detector exposure
and beam, assuming the normal mass hierarchy and $\delta = 0$.
The background rate
for the anti-neutrino beam is smaller than
that for the neutrino beam by a factor of 1.6. 
Since the signal rate for the anti-neutrino beam is
smaller by a factor of 3.4, the signal to noise ratio is
worse for the anti-neutrino beam than that for the neutrino beam
by a factor of about 2.

We assume that the experiment has a front detector that
measures the background and the signal detection efficiency.
We assume that both the overall background rate and the 
background shape (i.e., the energy dependence of the background)  
are understood within the uncertainty of 5\%. More details
on the assumed uncertainty in the background will be discussed 
later in this subsection when  
we define the $\chi^2$ for the analysis.  
We stress that in the present setting 
the detectors located in Kamioka and in Korea are not only 
identical but also they receive neutrino beams with the same energy 
distribution (due to the same off-axis angle) in the absence of oscillations. 
Therefore, the dependence on the assumed value of the experimental 
systematic errors must be weak, as we will verify by repeating 
the same analysis with different values of the uncertainties. 
As we will see later, the error cancellation between identical detectors 
gives us a very powerful constraint in our analysis.

We compute neutrino oscillation probabilities 
by numerically integrating neutrino evolution equation 
under the constant density approximation.  
The average density is assumed to be 2.3 and 2.8 g/cm$^3$ for the 
matter along the beam line between the production target and 
Kamioka and between the target and Korea, respectively.\footnote{
The former number, as  quoted in \cite{koike-sato},  is the average 
value of the matter density profile  which is obtained by M. Komazawa at 
Geographical Survey Institute in Japan by measuring surface gravity 
in many points around the path from KEK to Kamioka for the K2K 
experiment. 
The latter one is a commonly taken value for the matter density 
in earth crust, which is roughly equal to the average density of 
the distribution expected by ``Preliminary Reference Earth Model'' 
\cite{Dziewonski}. 
}
We assume that the number of electron with respect to that of nucleons 
to be 0.5 to convert the matter density to the electron number density.

\begin{figure}[htbp]
\vglue 0.3cm
\begin{center}
\includegraphics[width=0.8\textwidth]{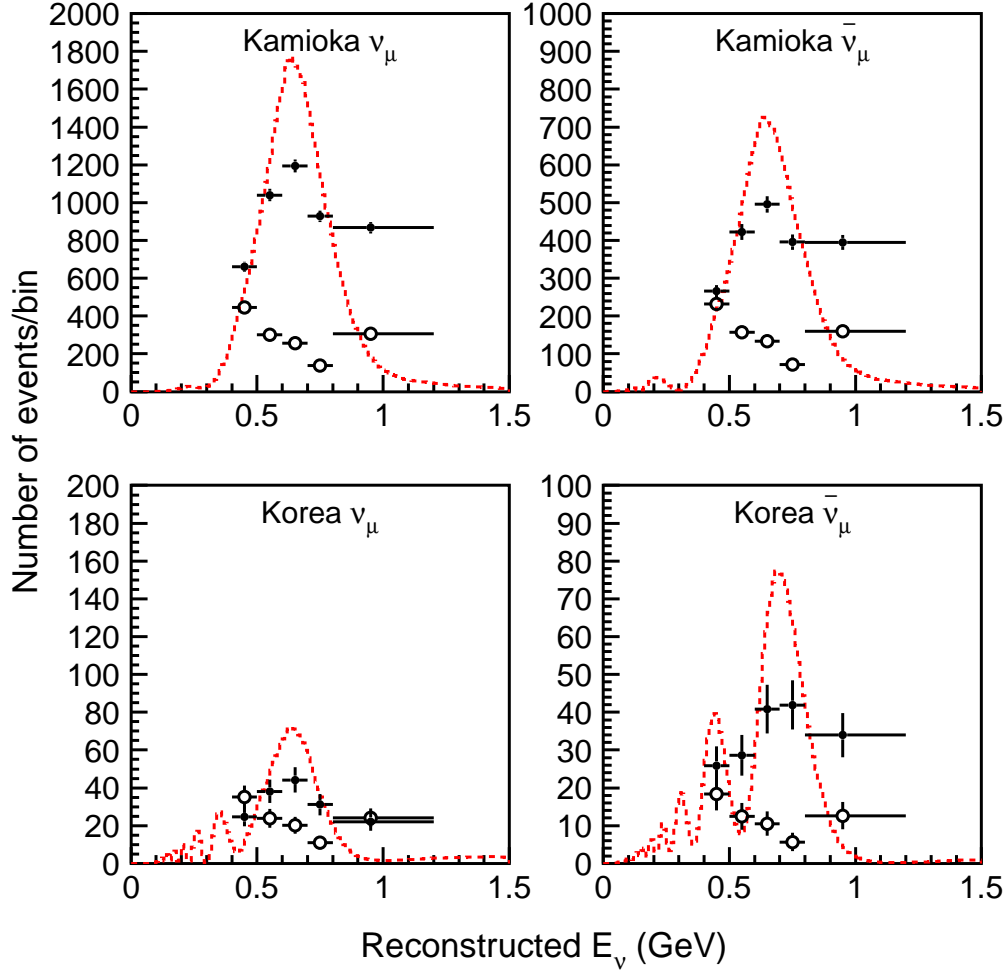}
\end{center}
\vglue -0.3cm
\caption{Examples of electron events to be observed in Kamioka
and in Korea for 4 years of operation of the neutrino and anti-neutrino beams 
are presented as a function of reconstructed neutrino energy. 
The fiducial masses are taken to be 0.27~Mton for both the detectors in 
Kamioka and in Korea.
The open circles show the background events. 
The solid circles and dashed lines show the expected
energy spectrum of signal events with and without the detector 
effects, respectively.    
$\sin^2 2 \theta_{13} =$0.1, $\delta = \pi/2$ and normal mass
hierarchy are assumed in simulating the events. 
}
\label{fig:energy-spectrum-examples}
\end{figure}

Fig.~\ref{fig:energy-spectrum-examples} shows an
example of the energy spectrum of electron events
to be observed in Kamioka and Korea for 4 years of
neutrino beam and 4 years of anti-neutrino beam.
Even with the 4~MW proton beam and with 0.27~Mton 
fiducial volume, the number of expected events in Korea is
limited due to the large distance between the production point and 
the detector. However, it is seen that the overall signal rate
and the energy dependence of 
the signal events are different between Kamioka and Korea.
Therefore, we use these features to obtain information on
neutrino oscillation parameters.

The statistical significance of the measurement considered in this
paper was estimated by using the following 
definition of $\chi^2$ \cite{pull}:
\begin{eqnarray}
\chi^2 =  \sum_{k=1}^{4} \left( \sum_{i=1}^{5}
\frac{\left(N(e)_{i}^{\rm obs} - N(e)_{i}^{\rm exp}\right)^2}
{ \sigma^2_{i} } \right)
+ \sum_{j=1}^{3} \left(\frac{\epsilon_j}{\tilde{\sigma}_{j}}\right)^2
\label{equation:chi2def}
\\
 N(e)_{i}^{\rm exp} = N_{i}^{\rm BG} \cdot 
   (1+\sum_{j=1}^{2}f_{j}^{i}\cdot\epsilon_{j}) 
    + N_i^{\rm signal} \cdot 
   (1+f_{3}^{i}\cdot\epsilon_{3}) ~~.
\label{equation:number}
\end{eqnarray} \noindent
%
The first sums are for the number of observed 
single-ring electron events in the $i^{\rm th}$ energy bin,
$N(e)^{\rm obs}_i$ is the number of events to be observed for the
given oscillation parameter set,
 and $N(e)^{\rm exp}_i$ is the expected number of
events for the assumed $\sin^2 2\theta_{13}$, $\delta$ and the mass 
hierarchy in the $\chi^2$ analysis, 
where $k=1,2,3$ and $4$ correspond to the four combinations 
of the detectors in Kamioka and in Korea with the 
neutrino and anti-neutrino beams,
respectively.
Both $N(e)^{\rm obs}_i$ and  $N(e)^{\rm exp}_i$ include
background events.
The energy ranges of the five energy bins are respectively
400-500~MeV, 
500-600~MeV, 600-700~MeV, 700-800~MeV and 
800-1200~MeV. 
$\sigma_i$ denotes the
statistical uncertainties in the expected data.
The energy resolution of 80 MeV is taken into account with a Gaussian 
resolution function. 

During the fit, the values of $N^{\rm exp}_i$ are
recalculated to account for neutrino oscillations, and so are the 
systematic variations in the predicted rates due to the uncertainties in the
estimated signal and background.  The overall background normalization 
is assumed to be uncertain by $\pm$5\%, and the effect is taken into 
account through $\epsilon_1$ in (\ref{equation:chi2def}).
In addition, it is also assumed that the background has an energy dependent 
uncertainty with the functional form of 
$((E_\nu(rec)-800~\text{MeV}) / 400~\text{MeV})\times (1 + \epsilon_2$). 
The energy-dependent part is also assumed to be uncertain by 5\%.
The uncertainty in the detection efficiency of the electron and positron 
signals is assumed to be 5\%.
In summary, 
$\tilde{\sigma}_{j} = 0.05$ for $j=1,2$, and 3. 

$N^{BG}_i$ is the number of background events
for the $i^{\rm th}$ bin.
We note that the number of the background events at detectors in 
Kamioka and in Korea are related simply by
$(L_{Korea}/L_{Kamioka})^2$, where $L$ is the
distance between the neutrino production point and the
detector.
$N^{signal}_i$ is the number of events
that are appeared by neutrino oscillations, and depend on 
$\sin^2 2\theta_{13}$, $\delta$ and the mass hierarchy.
The uncertainties in $N^{BG}_i$ and $N^{signal}_i$
are represented by 3 parameters $\epsilon_j$. 
The parameter $f^i_j$ represents the fractional change in the predicted
event rate in the $i^{\rm th}$ bin due to a variation of the parameter
$\epsilon_j$. The third sum in the $\chi^2$ definition collects the
contributions from variables which parameterize the systematic
uncertainties in the expected number of background events. 
 During the fit, these 3
parameters are varied to minimize $\chi^2$ for each choice of
the oscillation parameters.

\subsection{Sensitivity achievable by two-detector complex}

We have shown in Fig.~\ref{4fold} typical results of the sensitivity 
analysis by taking as an example two values of $\theta_{13}$, 
$\sin^2 2 \theta_{13} = 0.005$ and 0.02 with normal mass hierarchy as inputs. 
In the upper and the lower figures the cases of  
a single 0.54~Mton detector in Kamioka and 
two detectors of 0.27~Mton each in Kamioka and in Korea, respectively,
are examined. 
Now, we proceed to the sensitivity analysis of mass hierarchy 
and CP violation, which covers the entire region of the relevant 
parameter space.

We examine five different detector configurations, namely 
5 cases of volume ratios of Kamioka to Korean detectors, 
keeping the total volume constant; 
(a) a single 0.54~Mton fiducial mass detector 
(the total mass of which should be about 1~Mton) in Kamioka,
(b) 0.27~Mton detectors in Kamioka and in Korea, 
(c) a 0.16~Mton detector in Kamioka and a 0.38~Mton detector in Korea 
(volume ratio of  3:7), 
(d) 0.05~Mton detector in Kamioka and a 0.49~Mton detector in Korea 
(volume ratio of  1:9),  and 
(e) a single 0.54~Mton detector in Korea.
It is assumed that the experiment will continue for 8 years
with 4 year of neutrino beam and 4 years of anti-neutrino beam.
We use the same definition of $\chi^2$ as in (\ref{equation:chi2def})
which embodies cancellation of errors between identical detectors 
even in the cases of two detectors with different volumes. 
For this simplification 
and the other reasons our analysis cannot be considered as 
a real optimization procedure for the volume ratio, but we feel that 
such study may give some insights on how the sensitivity varies with the 
detector mass ratio.

\begin{figure}[htbp]
\vglue 0.3cm
\begin{center}
\includegraphics[width=0.73\textwidth]{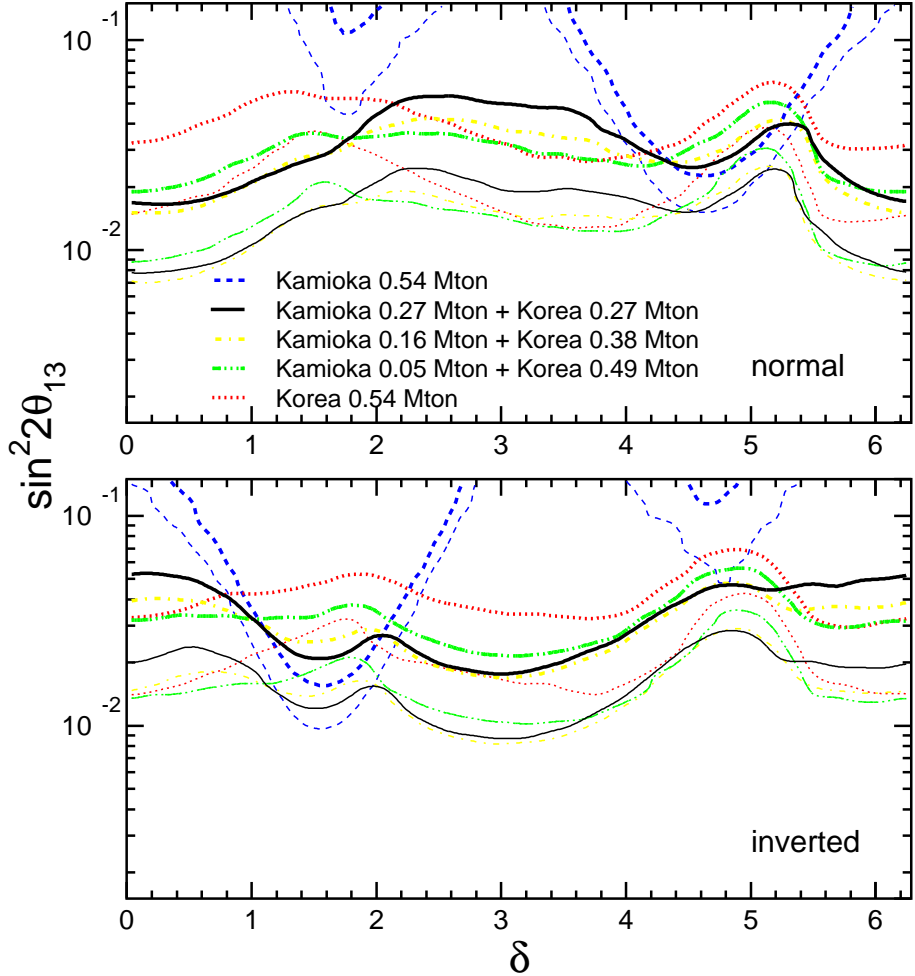}
\end{center}
\vglue -0.3cm
\caption{ 2(thin lines) and 3(thick lines) 
standard deviation sensitivities to  the mass hierarchy for 
the volume ratio of Kamioka to Korea 1:0 
(0.54~Mton fiducial mass detector in Kamioka, indicated by dashed lines, blue),
the volume ratio 1:1 
(0.27~Mton detectors in Kamioka and in Korea, solid lines, black), 
the volume ratio 3:7 
(a 0.16~Mton detector in Kamioka and a 0.38~Mton detector in Korea, 
dash-dot lines, yellow), 
the volume ratio 1:9
(a 0.05~Mton detector in Kamioka and a 0.49~Mton detector in Korea, 
dash-dot-dot lines, green), 
and the volume ratio 0:1 
(a 0.54~Mton detector at Korea, dotted lines, red).
4 years running with neutrino beam and another 4 years with 
anti-neutrino beam are assumed.
}
\label{sensitivity-detectors_hierarchy}
\end{figure}

We show in 
Fig.~\ref{sensitivity-detectors_hierarchy} 
the contours of sensitivity to the mass hierarchy 
at 2 and 3 standard deviations on $\delta$-$\sin ^2 2 \theta_{13}$ 
plane for these 5 detector configurations. Here, the 
2 and 3 standard deviations are defined to be 
$|\chi^2_{min} \text{(wrong~hierarchy)}-\chi^2_{min}\text{(true~hierarchy)}|>$
4 and 9, respectively. 
In the upper and the lower panels of Fig.~\ref{sensitivity-detectors_hierarchy},  
cases of the normal and the inverted mass hierarchies as a true mass 
hierarchies are presented. 
As expected, better sensitivities are obtained for detector configurations 
with detectors both in Kamioka and in Korea. 
While the best sensitivity is obtained for the case of (c) 3:7 volume ratio, 
we find that the sensitivity on the mass hierarchy does not depend 
strongly on the mass ratio of the two detectors.
It is interesting to note that there are some sensitivities to mass hierarchy 
in a limited region of $\delta$ even in the setting with a single detector in 
Kamioka, as expected \cite{MNjhep01,nufact01_mina}.

\begin{figure}[htbp]
\vglue 0.3cm
\begin{center}
\includegraphics[width=0.73\textwidth]{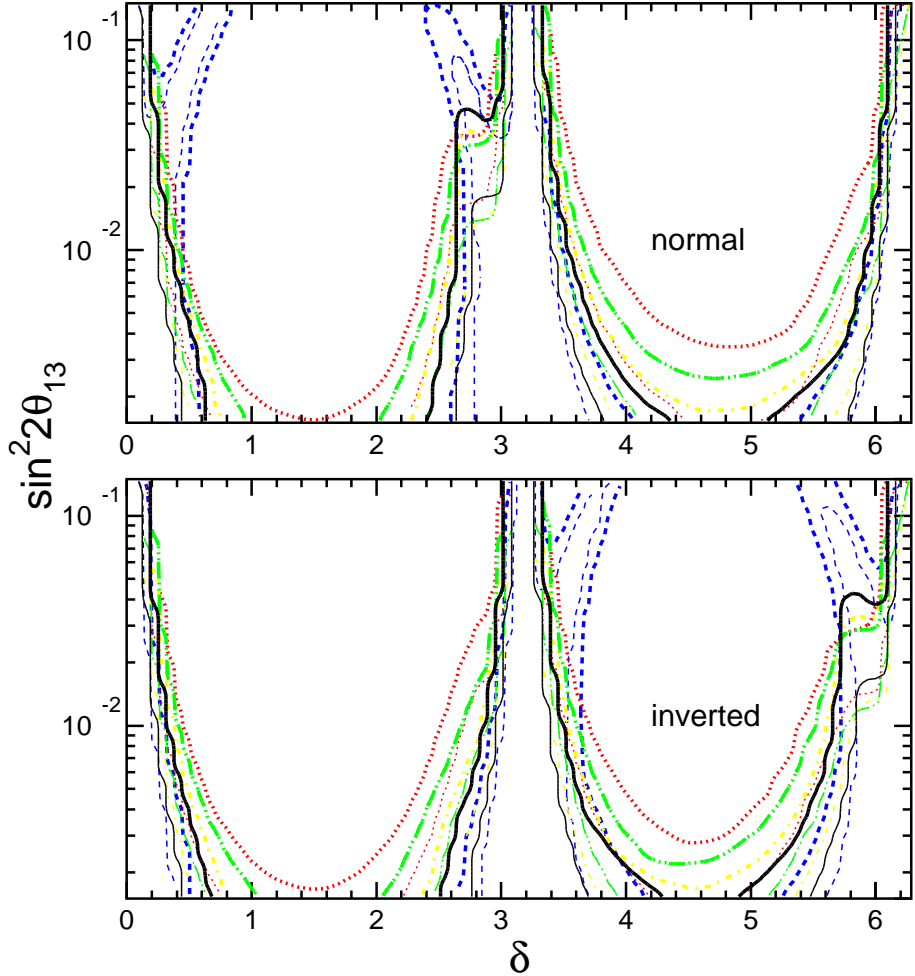}
\end{center}
\vglue -0.3cm
\caption{ The same as in Fig.~\ref{sensitivity-detectors_hierarchy} but for the sensitivity to leptonic CP violation. 
The correspondence between line symbols and detector configurations 
is the same as in Fig.~\ref{sensitivity-detectors_hierarchy}. 
}
\label{sensitivity-detectors_CP}
\end{figure}

We show in 
Fig.~\ref{sensitivity-detectors_CP} 
the sensitivity to leptonic CP violation by drawing the contours 
above which hypothesis of $\delta = 0$ or $\pi$ can be rejected at
2 and 3 standard deviations. 
The 2 and 3 standard deviations are defined to be 
$|\chi^2_{min}(\delta\not=0~\text{or}~\pi) - \chi^2_{min}(\text{true value of}~\delta)| >$
4 and 9, respectively. 
As in the upper and the lower panels of 
Fig.~\ref{sensitivity-detectors_hierarchy}, 
the cases of the normal and the inverted mass hierarchies as a 
true mass hierarchies are presented. 
Notice that while the type of true hierarchy is assumed in each case, 
the fit is performed without assuming the mass hierarchy. 
This statement applies to all the figures presented in this section. 
For small $\sin ^2 2 \theta_{13}$ a single 0.54~Mton detector in Kamioka
has the best sensitivity. Whereas for large $\sin^2 2 \theta_{13}$, 
the two detector configuration gives the better sensitivity.
It is due to the fact that in the two detector configurations, the 
mass hierarchy can be resolved for large $\sin ^2 2 \theta_{13}$, 
and therefore the value of $\delta$
is uniquely measured in the case of the two detector configurations.
Among the two detector configurations, 0.27~Mton detectors at
Kamioka and Korea gives the best sensitivity to the CP violation
measurement.

We observe that the sensitivities to the mass hierarchy and CP violation 
do not change appreciably as the volume ratio is varied. 
A larger Korean detector seems to be favored to resolve 
the mass hierarchy. 
On the other hand, the sensitivity to CP violation disfavors the larger 
Korean detector, and in fact the best choice would be a 0.54 Mton 
detector in Kamioka for small $\sin ^2 2 \theta_{13}$ values.
Overall,  two 0.27 Mt detectors one in Kamioka and the other in Korea 
seems to be close to the optimal choice. 
Anyhow  the difference is so small that the real optimization process 
must involve other factors, such as the site availability and conditions, 
as well as additional physics capabilities, etc.

We note, in passing, that in the case of a single 0.54 Mton 
detector in Kamioka the spectral analysis is so powerful that 
allows larger sensitivity region of discovering CP violation 
than the one presented in \cite{JPARC} 
(which relied on rate only analysis), as one can see in 
Fig.~\ref{sensitivity-detectors_CP}. 
It is highly nontrivial because we used relaxed values of systematic errors of 5\%.
Moreover, one can recognize the power of the spectrum information 
to solve the intrinsic degeneracy.
%

>From the results obtained in this subsection, we select out 
the two identical detector configuration with 0.27~Mton each in 
Kamioka and in Korea for further studies of the 
sensitivities of resolving the mass hierarchy and the CP degeneracy, 
which is to be carried out in the following subsections.

\subsection{Optimization of neutrino and anti-neutrino running time}

\begin{figure}[htbp]
\begin{center}
\includegraphics[width=0.58\textwidth]{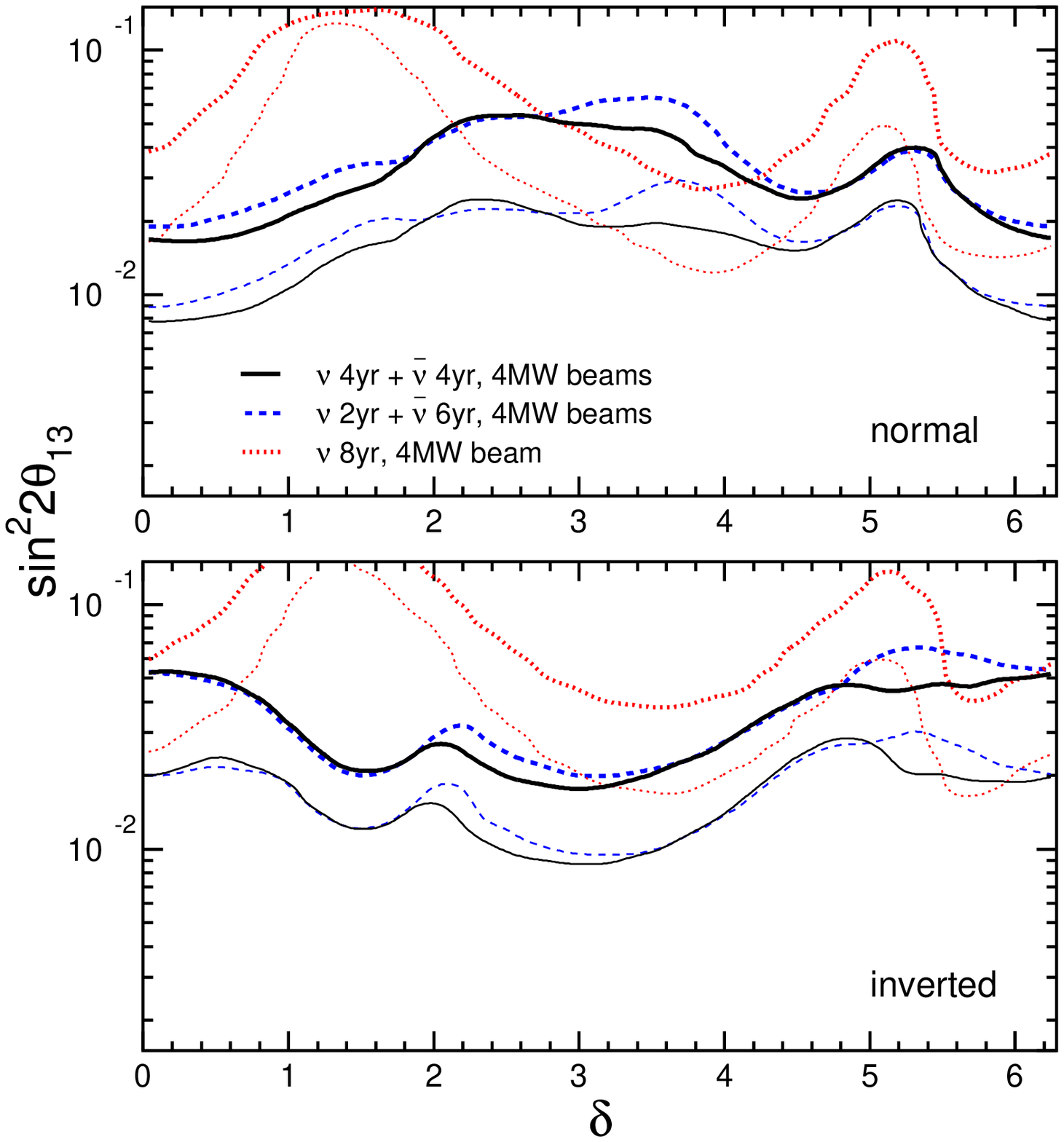}
\includegraphics[width=0.58\textwidth]{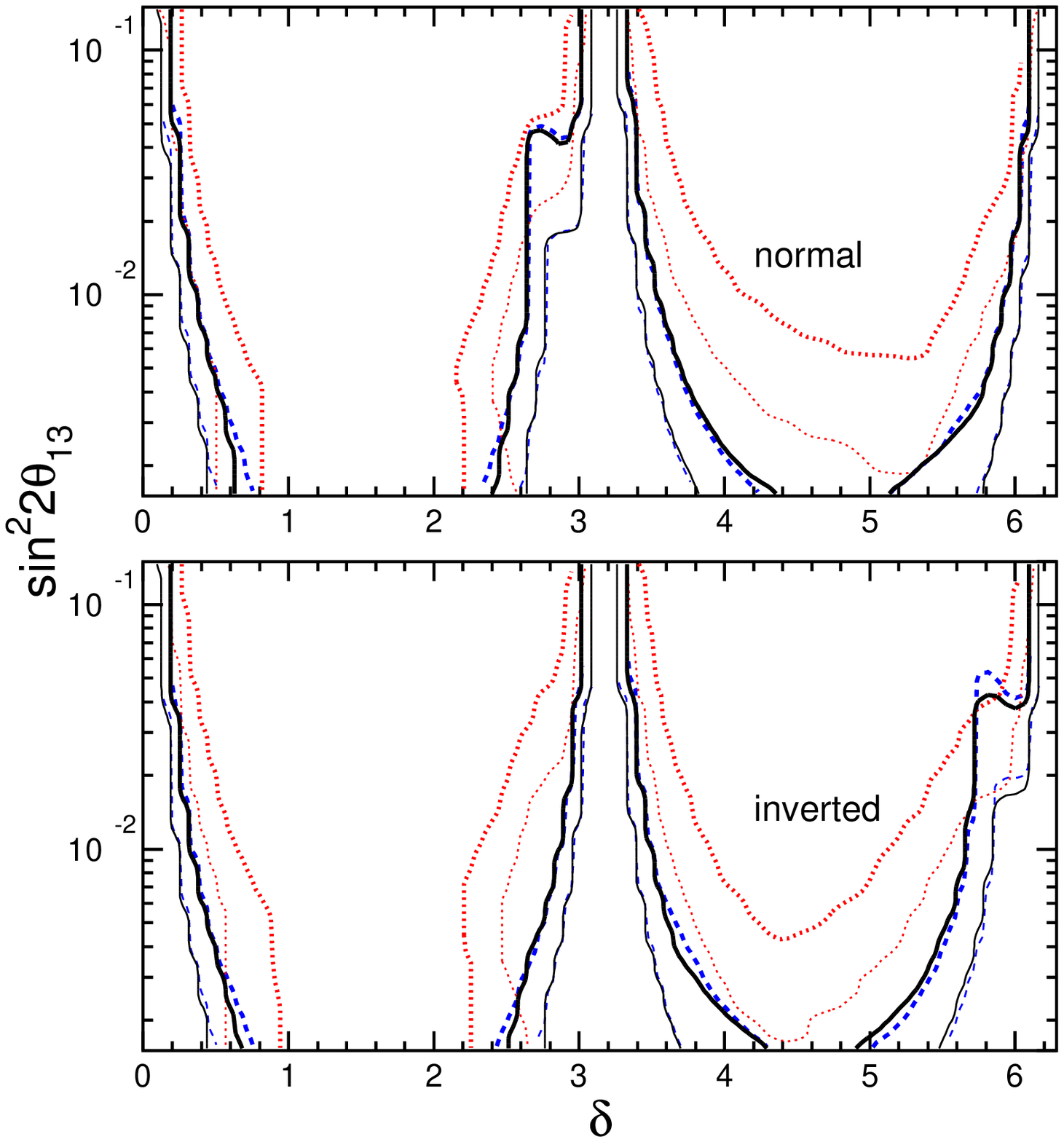}
\end{center}
\vglue -0.3cm
\caption{ 2(thin lines) and 3(thick lines) 
standard deviation sensitivities to; the mass hierarchy (upper panels)  
and the leptonic CP violation (lower panels) for
0.27~Mton detectors in Kamioka and in Korea
with 4 years running with neutrino beam and 4 years with 
anti-neutrino beam (solid lines, black),
8 years with neutrino beam only (dotted lines, red) and  
2 years with neutrino beam and 6 years with anti-neutrino beam 
(dashed lines, blue).
}
\label{sensitivity-beams}
\end{figure}

Since we found that the best detector configuration is the 
identical detectors in Kamioka and in Korea, we search
for the best option for dividing assumed whole period of 8 years 
into the neutrino and anti-neutrino running. 
We examine the following three cases: 
(i) 8 years of the neutrino beam, 
(ii) 4 years of the neutrino and anti-neutrino beams (4+4), and  
(iii) 2 years of the neutrino beam and 6 years of the 
anti-neutrino beam (2+6).
The last option is studied
in Ref.~\cite{JPARC} where the sensitivity on the CP violation is
discussed with a single detector in Kamioka. Figure~\ref{sensitivity-beams}
(upper panels) shows the sensitivity to the mass hierarchy 
 in $\delta$-$\sin ^2 2 \theta_{13}$ space at 2 and 3 standard
deviations for the three beam options. We find that the (4+4) 
option gives the best sensitivity while the (2+6) option gives
similar but somewhat lower sensitivity. 

Figure~\ref{sensitivity-beams} (lower panels) 
shows the sensitivity to the leptonic 
CP violation in $\delta$-$\sin ^2 2 \theta_{13}$ space at
 2 and 3 standard deviations for the three beam options. 
The (4+4) and (2+6) options have almost the same sensitivity to the 
leptonic CP violation. 
We note that this study suggests that the leptonic CP violation can be 
demonstrated by running the experiment with the neutrino beam alone 
for some values of $\sin^2 2 \theta_{13}$ and $\delta$ 
though the sensitivity is substantially lower than the neutrino plus 
anti-neutrino beam options. This can be understood because
the energy spectrum data from the detectors in Kamioka
and Korea contain information on $\delta$.

%
\begin{figure}[htbp]
\vglue 0.3cm
\begin{center}
\includegraphics[width=0.46\textwidth]{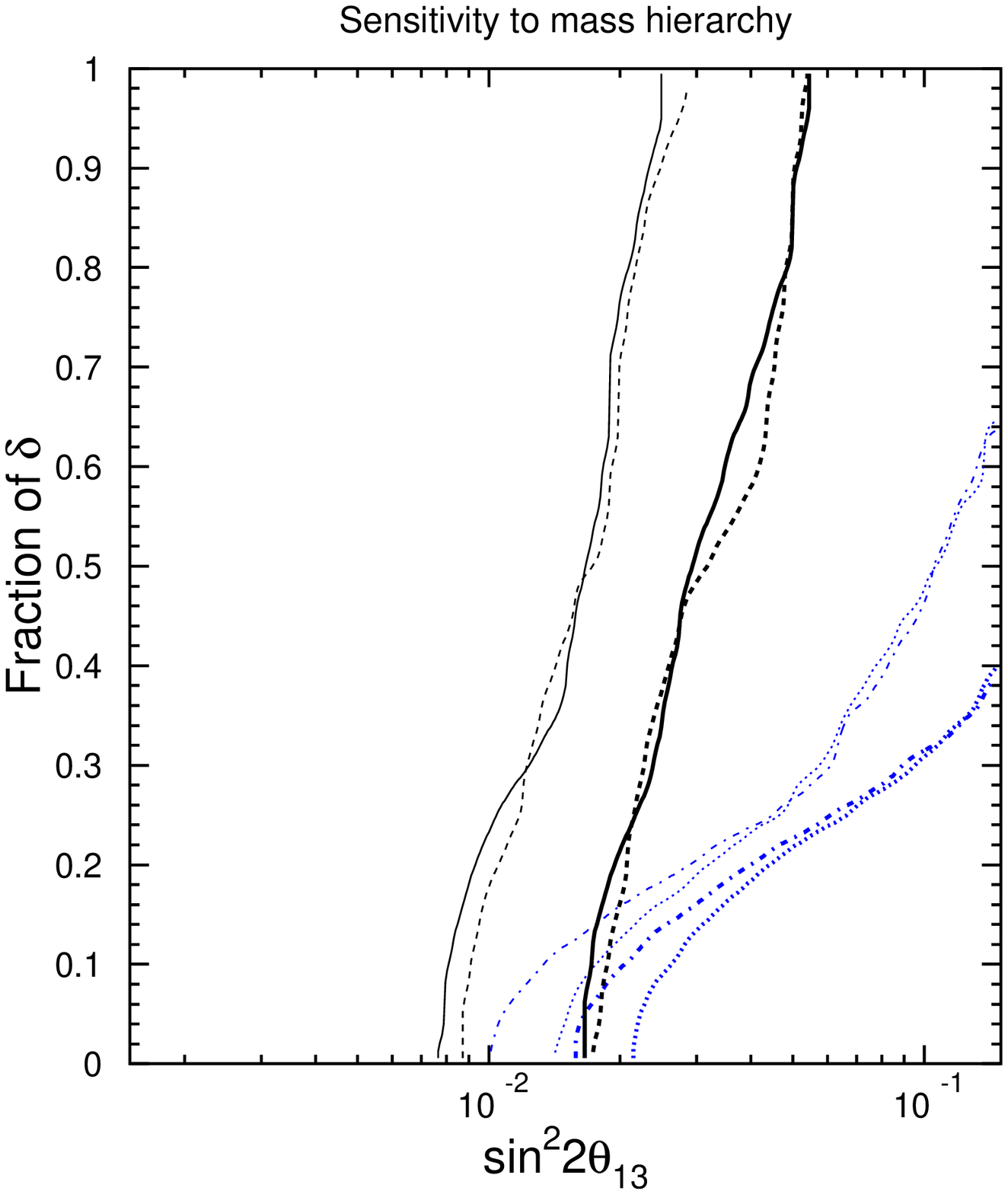}
\includegraphics[width=0.46\textwidth]{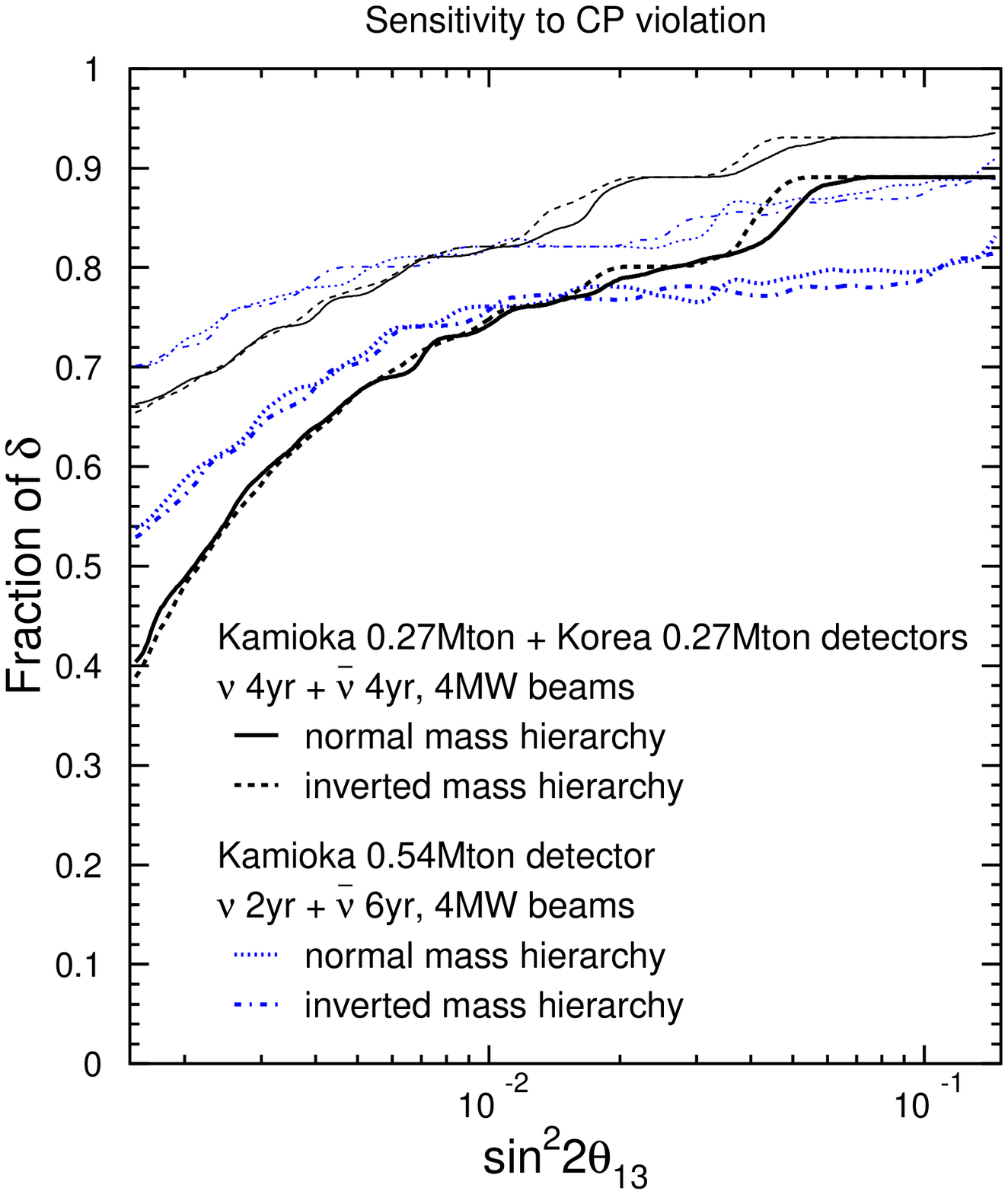}
\end{center}
\vglue -0.3cm
\caption{ 2(thin line) and 3(thick lines) 
standard deviation sensitivities to mass hierarchy 
(left panel) and CP violation (right panel). Solid and dashed black lines
show the cases of two identical detectors with 
fiducial mass of 0.27 Mton in Kamioka and in 
Korea with 4 years of neutrino and 4 years of anti-neutrino beam running. 
Dotted and dash-dotted blue lines show the cases for a single detector with 
fiducial mass of 0.54 Mton in Kamioka 
with 2 years of neutrino and 6 years of anti-neutrino beams.
Solid (dotted) and dashed (dash-dotted) lines show the cases for positive and 
negative mass hierarchies, respectively.
}
\label{fraction}
\end{figure}

In Fig.~\ref{fraction} we present sensitivity regions for the 
mass hierarchy and CP violation at 2 
and 3 standard deviations 
in space spanned by $\sin^2 2\theta_{13}$ and 
the fraction of $\delta$. 
The plot probably needs explanation: 
For a given value of $\sin ^2 2\theta_{13}$, one can draw the region, 
as we did in Fig.~\ref{sensitivity-detectors_hierarchy},  
Fig.~\ref{sensitivity-detectors_CP},  and Fig.~\ref{sensitivity-beams} 
in the previous subsections,  where the mass hierarchy can 
be resolved (or CP violation can be signaled) at 2 or 
3 standard deviations. The plot can be converted to the one in 
Fig.~\ref{fraction} by computing the fraction of the region of $\delta$ 
above the resolution contour at the given value of $\sin ^2 2\theta_{13}$. 
In this way, the plot gives an alternative way of representing 
how sensitive is the experiment to the mass hierarchy or CP violation, 
and is used e.g., in the proposal of the NO$\nu$A experiment 
\cite{NOVA}. 

In the proposed experiment, the mass hierarchy can be resolved if
$\sin ^2 2 \theta_{13}$ is larger than 0.055 (0.03) at the 3 (2)  
standard deviation level for any $\delta $ values. 
(Note, however, that the sensitivity is better in most of the region of 
$\delta $ as can be seen in Fig.~\ref{sensitivity-detectors_hierarchy}.)

\subsection{Systematic uncertainties and the sensitivity}

\begin{figure}[htbp]
\begin{center}
\includegraphics[height=7.9cm]{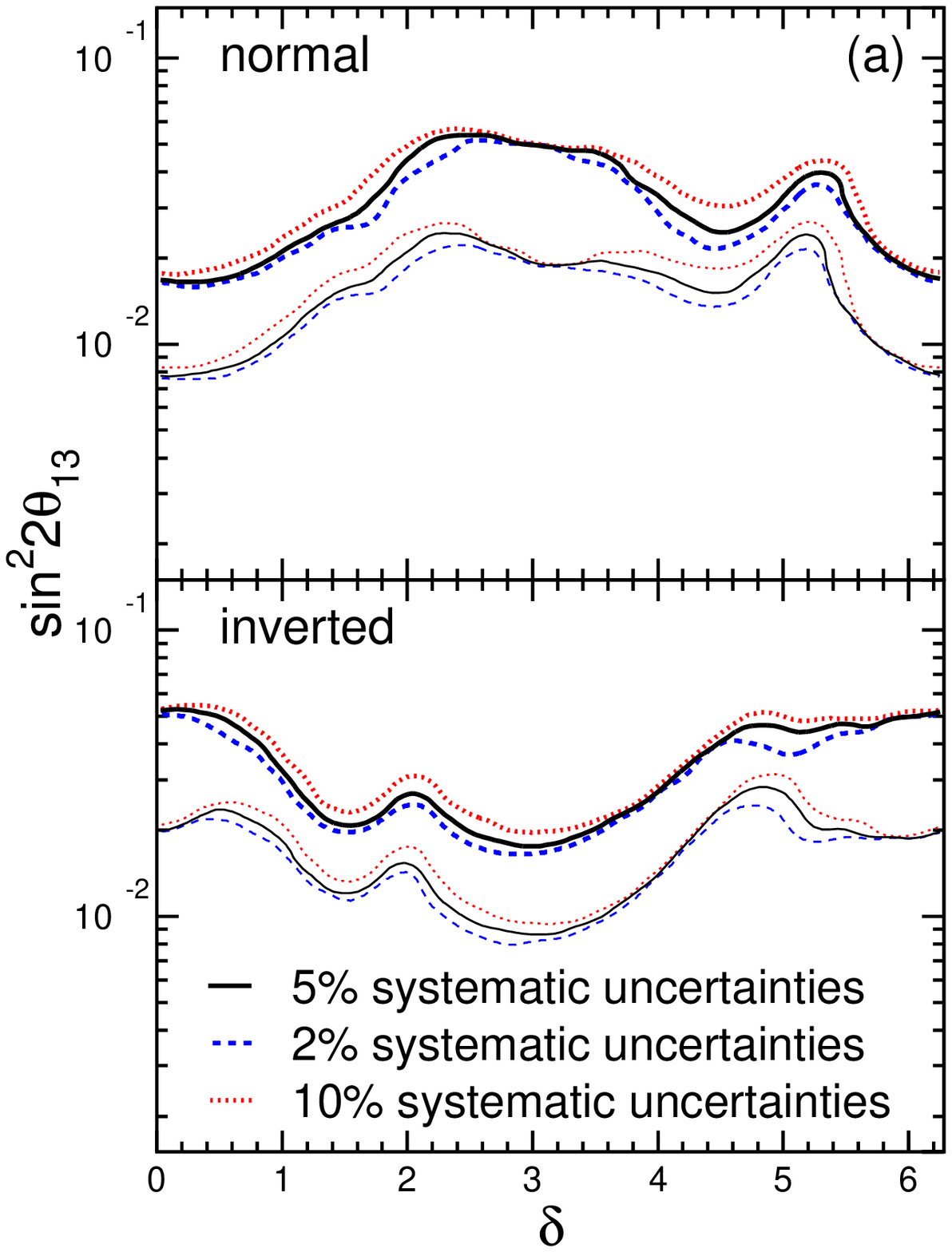}
\includegraphics[height=7.9cm]{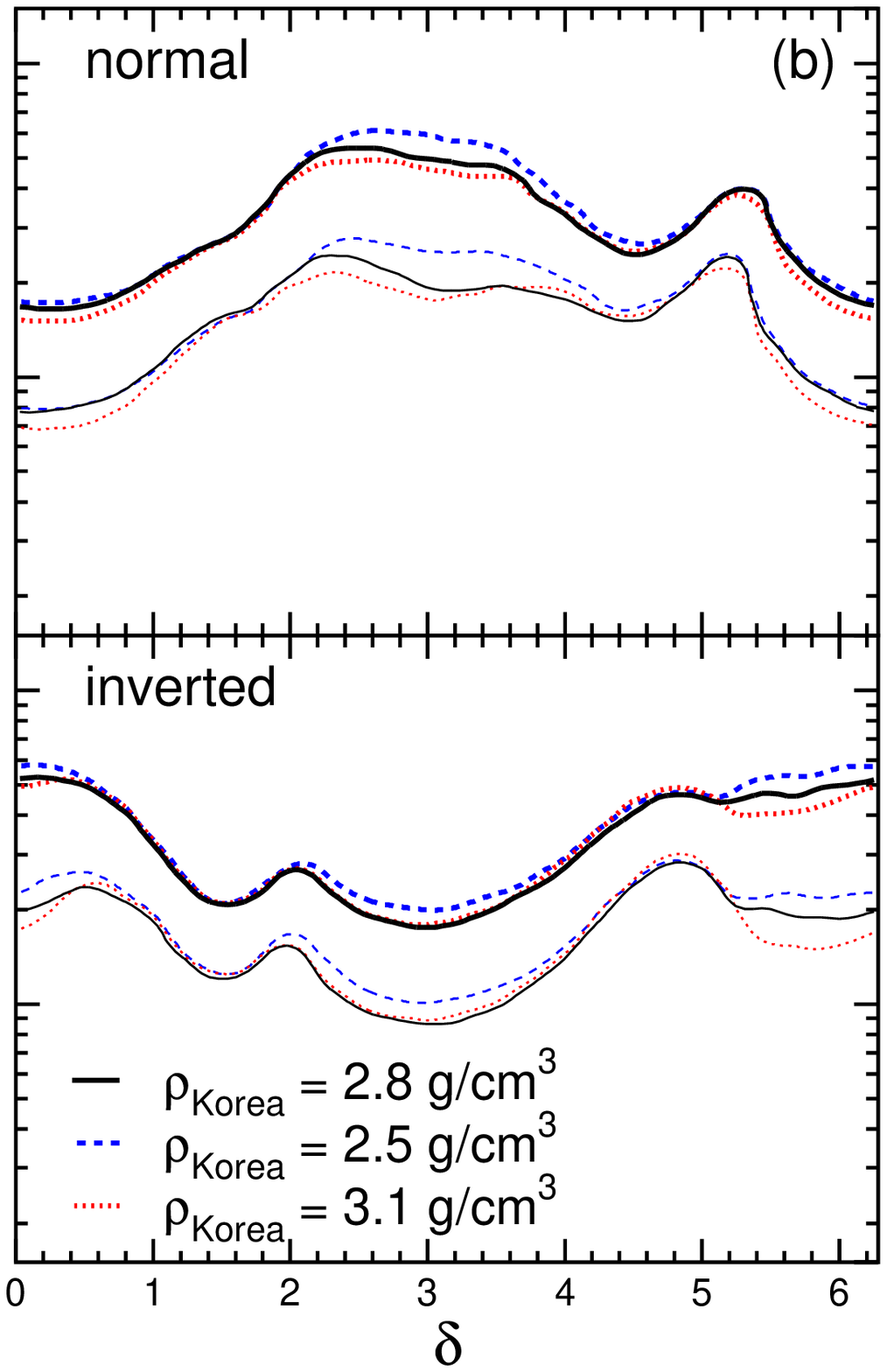}
\includegraphics[height=7.9cm]{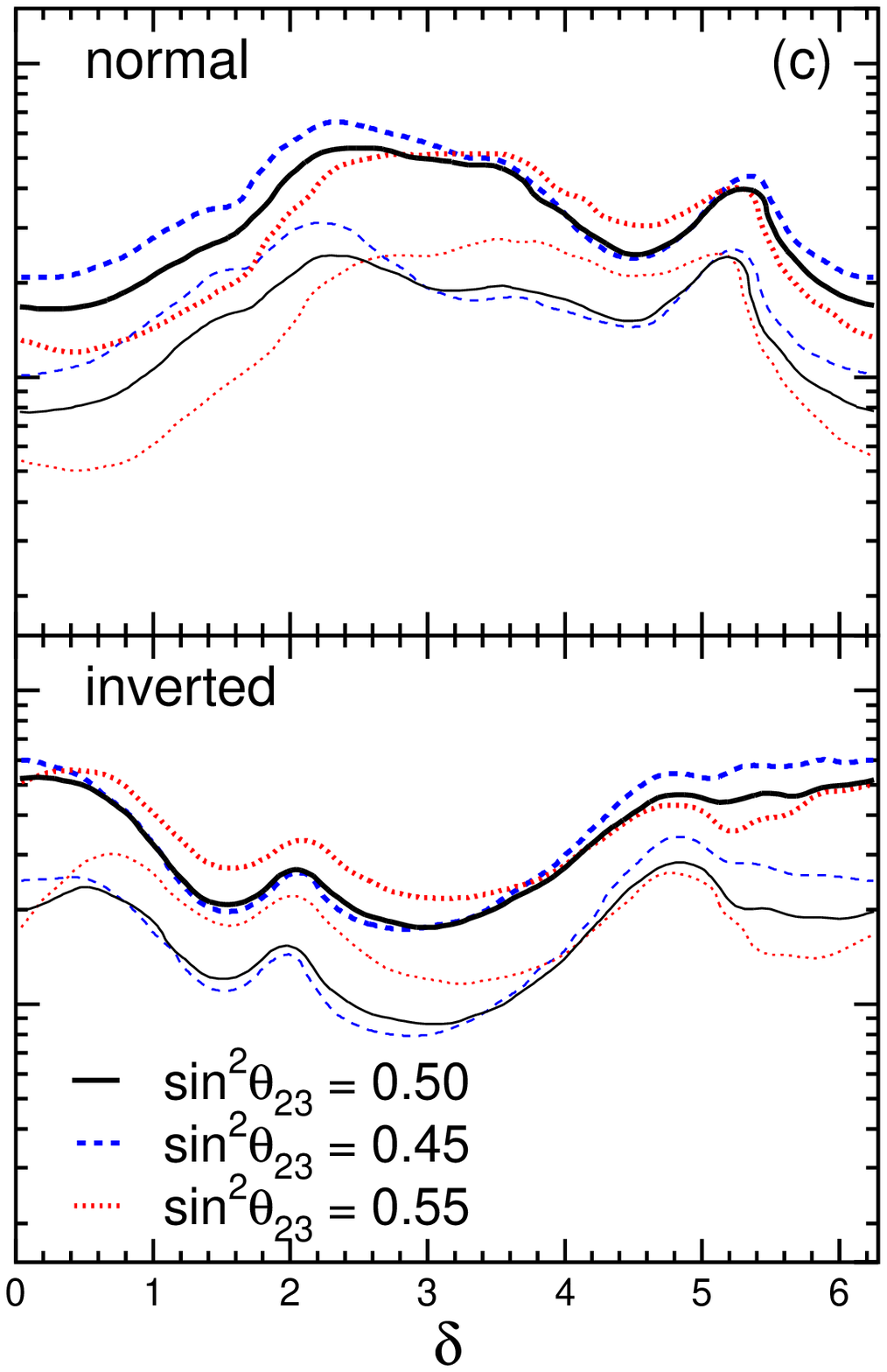}
\end{center}
\vglue -0.5cm
\caption{ (a) (left panels): 2(thin line) and 3(thick lines) 
standard deviation sensitivities to mass hierarchy for systematic uncertainties of 
2\% (dashed lines, blue),
5\% (solid lines, black), and 
10\% (dotted lines, red) 
in the estimated signal and background. 
(b) (middle panels): the same as in (a) but with varying true values 
of mean matter density,   
2.5 (dashed lines, blue), 2.8 (solid lines, black), and 
3.1\,$\text{g/cm}^3$ (dotted lines, red). 
The analysis is carried out assuming
the matter density of 2.8 $\text{g/cm}^3$. 
(c) (right panels): the same as in (a) but with varying true values of 
$\sin ^2 \theta_{23}$, 0.45 (dashed lines, blue), 
0.5 (solid lines, black), and 0.55 (dotted lines, red).  
The analysis is carried out assuming 
$\sin ^2 \theta_{23} = 0.5$
}
\label{check_systematic-error}
\end{figure}

In order to study the robustness of the results, we carry 
out several tests. First, we test stability of the results 
by varying the systematic uncertainties in the background
estimation and the signal detection efficiency. We examine 
three values, 2, 5, and 10\%. 
Figure~\ref{check_systematic-error}a
 shows the sensitivity region to the mass hierarchy
in $\delta$-$\sin ^2 2 \theta_{13}$ space at 2 and 3 standard
deviations for the three values of systematic uncertainties. 
It is remarkable that the dependence of the sensitivity 
to the systematic error is extremely weak. 
We point out that, because of the identical energy spectrum and the 
detectors in Kamioka and in Korea, their background rates must be 
related simply by $(L_{\text{Korea}} / L_{\text{Kamioka}})^2$. 
Because of this relation, the difference in the signal events in Kamioka 
and Korea cannot be explained by the uncertainty in the background, 
and therefore the measurement of the background rate by front 
detectors with a very high precision is not crucial in the identical 
two detector setup.

%
The matter density along the neutrino beam is not precisely known. 
Uncertainty in the matter density between 
J-PARC and Korea could be non-negligible, 
while the matter effect could be more important.
We study the systematic effect due to the
uncertainty in the matter density by changing only the mean 
matter density between the target and Korea. 
The true mean matter density is assumed to be either 2.5, 2.8 or 3.1\,g/cm$^3$,
while the analysis is carried out assuming the density of 2.8\,g/cm$^3$. 
Figure~\ref{check_systematic-error}b (middle panel)
 shows contours of the sensitivity to mass hierarchy in 
$\delta$-$\sin ^2 2 \theta_{13}$ space for three cases of the matter density. 
As indicated in the figure, there is only a small difference in the 
sensitivity to the mass hierarchy.

Even if $\sin ^2 2 \theta_{23}$ is measured to an accuracy of
1\%, the uncertainty in $\sin^2 \theta_{23}$ is relatively large.
To examine the dependence of the sensitivity to $\theta_{23}$, we 
have assumed that the true value of $\sin ^2 \theta_{23}$ is 
either 0.45, 0.5 or 0.55,
while the analysis is carried out assuming $\sin ^2 \theta_{23} =$0.5. 
Figure~\ref{check_systematic-error}c (right panel) 
shows the sensitivity to the mass hierarchy in 
$\delta$-$\sin ^2 2 \theta_{13}$ space for the three values of 
$\sin ^2 2 \theta_{23}$.
Although the sensitivity to mass hierarchy is somewhat influenced by 
the uncertainty of $\sin ^2 \theta_{23}$, the mass hierarchy can still be 
resolved without any significant overall change in the sensitivity.
The effect of the uncertainty in $\sin^2 \theta_{23}$
 to the sensitivity to the leptonic CP violation is very small.

In summary, we find that the conclusion we presented in the previous
subsection does not depend strongly on the assumptions we have made.

\section{Concluding remarks}
\label{conclusion}

In this paper, we have explored physics capabilities of the 
two identical megaton-class detector complex (twin HK), 
one placed in the Kamioka mine and the other in Korea 
under the assumption that they receive the same neutrino beam 
from J-PARC with 4MW beam power.  
Unlike the foregoing analyses of such two detector setup, 
we have employed a radical new strategy of resolving the 
$\Delta m^2_{31}$-sign degeneracy 
to determine simultaneously the neutrino mass hierarchy and the 
CP violating phase $\delta$. 
We have demonstrated that the two-detector complex can determine 
neutrino mass hierarchy down to 
$\sin^2 2\theta_{13} \gsim 0.02$ (0.05) for $\simeq$80\% coverage of 
the whole range of $\delta$, and 
$\sin^2 2\theta_{13} \gsim 0.03$ (0.054) for any value of $\delta$ both 
at 2$\sigma$ (3$\sigma$) CL, as indicated in Fig.~\ref{fraction}. 
Importantly, the sensitivity to the CP violation of the current design 
of J-PARC phase II project is essentially kept or even enhanced 
at  $\sin^2 2\theta_{13} \gsim 0.01$. 
The key to the enormous sensitivities is the use of two identical detectors 
that allows significant reduction of systematic errors by cancellation. 
It allows unambiguous detection of spectral distortion of neutrino 
energy distribution due to oscillations, which is crucial to resolve the 
$\Delta m^2_{31}$-sign (the intrinsic) degeneracy with (without) 
help of the matter effect.

We have shown that our proposal elevates the current 
design of the phase II of J-PARC neutrino program which aims 
at discovering leptonic CP violation to a dual-purpose experiment, 
in which one can determine neutrino mass hierarchy and CP violation 
up to a level achievable by conventional superbeam experiments. 
We emphasize that, in our setting, the total volume of the two HK 
is the same as the planned HK with 1 Mton water in Kamioka, 
and the current design of the latter option already contains two tanks. 
Hence, the cost of the present proposal should be roughly the same as that 
of the conventional design. 
Yet, we neither addressed the issue of site availability in Korea, nor the 
precise estimate of systematic errors which requires specification of 
the site as well as actual design of the detectors. 

There exist several other proposals for the 
 next-generation long-baseline experiments in the world, 
NO$\nu$A \cite{NOVA}, the very-long-baseline project \cite{BNL}, 
and the beta beam \cite{beta}. 
With different assumptions and conditions, it 
is difficult to compare the present proposal  
with the existing ones.
Nonetheless, we would like to emphasize that the present proposal 
is at least competitive to them not only on the sensitivity to the 
CP violation 
but also on resolving power of the neutrino mass hierarchy.

\begin{acknowledgments}
One of the authors (TK) thanks Edward Witten for the encouragement
of exploring the possibility discussed in this paper. 
This work was supported in part by the Grant-in-Aid for Scientific Research, 
Nos. 15204016 and 16340078, Japan Society for the Promotion of Science
and by Conselho Nacional de Ci\^encia e  Tecnologia (CNPq). 
\end{acknowledgments}


\end{document}